\documentclass[11pt,a4paper]{article}
\usepackage[margin=28mm,headheight=14pt]{geometry}
\usepackage{amsmath,amssymb,amsthm,mathtools}
\usepackage{fontspec}

\usepackage{microtype}
\usepackage{booktabs,longtable,tabularx,array}
\usepackage{enumitem}
\usepackage{needspace}
\usepackage{placeins}
\usepackage{fancyhdr}
\usepackage{fvextra}
\usepackage{xcolor}
\usepackage{tikz}
\usepackage[authoryear,round]{natbib}
\setcitestyle{aysep={,},yysep={,},notesep={, }}
\usepackage[colorlinks=true,citecolor=blue,linkcolor=black,urlcolor=black,unicode,bookmarksdepth=3,bookmarksnumbered=true,bookmarksopen=false]{hyperref}
\usepackage[nameinlink,noabbrev]{cleveref}
\newtheorem{theorem}{Theorem}[section]
\newtheorem{lemma}[theorem]{Lemma}
\newtheorem{proposition}[theorem]{Proposition}
\newtheorem{corollary}[theorem]{Corollary}
\theoremstyle{definition}

\newcommand{\OUT}{\ensuremath{\mathrm{OUT}}}

\setlist{topsep=5pt,itemsep=3pt}
\title{Sequential Candidate Entry:\\ A Counterexample to the Osborne--Kats Conjecture}
\author{Dmitry Dagaev\textsuperscript{1,2}\\[0.4em]\small\textsuperscript{1}New Economic School\\\small\textsuperscript{2}HSE University}
\date{29 November 2026}
\begin{document}
\maketitle
\begin{abstract}
Finitely many players sequentially choose an electoral position on the interval $[0,1]$ or stay out, observing all prior choices. Voters have a uniform distribution of ideal positions on the same interval and vote non-strategically for a nearest candidate; candidates at the same position divide its votes. Candidates with the largest vote share divide a unit payoff equally, losing entrants receive a negative payoff, and staying out yields zero. In the mid-1980s, Osborne and Kats conjectured that, for every number $n\ge2$ of players, the unique subgame-perfect Nash equilibrium (SPNE) outcome has only the first and last players entering at the median. We refute the conjecture by establishing the existence of a different SPNE outcome for 12 players. Starting from the first two entries $1/2$ and $1/10$ proposed earlier by de Vries, we show that the resulting subgame admits an SPNE giving the second player a payoff of at least $1/10$.
\end{abstract}
\clearpage

\section{Introduction}\label{sec:introduction}
The threat of further entry is central to spatial electoral competition. A player who enters commits to a platform before later players decide whether and where to enter. The Osborne--Kats conjecture formulated in the 1980s makes a sharp prediction in this setting: regardless of the number of players, sequential entry yields the first and last players at the median, whereas all other players abstain \citep{osborne}.

We give a counterexample with 12 players and a uniformly distributed electorate. Following the first two entries $1/2$ and $1/10$ proposed earlier by \citet{devries}, we establish the existence of a subgame-perfect Nash equilibrium (SPNE) after these actions in which the second player receives a payoff of at least $1/10$. The proof selects full equilibrium strategy profiles in subsequent subgames and combines them with best responses. Some profiles and maximizing actions are chosen by existence arguments; we do not explicitly determine every contingent action or identify a particular equilibrium path. Extending suitably chosen subgame strategy profiles yields an SPNE of the full game with a noncanonical outcome. 

\label{sec:model}

Let $\Gamma$ denote the following game. There are 12 players, numbered in their order of decision. Each acts once and observes the complete preceding history. Write $a_i$ for player $i$'s action, a location in $[0,1]$ or $\OUT$. A \textit{history} $h=(a_1,\ldots,a_k)$ is a sequence of actions of the first
$k$ players, where $1\le k\le12$.  A \emph{candidate at $h$} is a player whose action in $h$ is a location. If $k<12$, player~$k+1$ moves next, and $(h,a)=(a_1,\ldots,a_k,a)$ denotes the history after that player chooses an available action $a$. Write $\Gamma(h)=\Gamma(a_1,\ldots,a_k)$
for the subgame following $h$, and $\Gamma(h,a)$ for the subgame
following $(h,a)$. At $k=12$, $\Gamma(h)$ is the trivial subgame with
no remaining moves. A \emph{terminal history} has length 12; a terminal extension
of $h$ is a terminal history beginning with $h$. The electorate has total mass one and is uniformly distributed on $[0,1]$. Each voter chooses a nearest candidate. Ties between distinct locations involve a set of voters of mass zero.
The \emph{vote cell} of a position chosen by at least 1 candidate
is the set of voter ideal points at least as close to that position
as to any candidate's position. Candidates at the same position divide the
mass of this cell equally. Let $y_1<\ldots<y_k$ be the distinct positions chosen by the candidates, and let $n_i$ be the number of candidates at $y_i$. When $k\geq2$, each candidate there receives vote share
\[
 v_i=\begin{cases}
 (y_1+y_2)/(2n_1),&i=1,\\
 (y_{i+1}-y_{i-1})/(2n_i),&1<i<k,\\
 (2-y_{k-1}-y_k)/(2n_k),&i=k.
 \end{cases}
\]
 
\noindent A player choosing $\OUT$ receives zero. Every
candidate with maximal vote share is called a \emph{winner}. Each
winner receives $1/m$, where $m$ is the number of winners; every
losing entrant receives $-1$. If everyone stays out, all payoffs are $0$.

These payoffs are equivalent to an office
prize $1$ and an entry cost $\kappa\in(0,1/12)$: winners receive
$1/m-\kappa$, losing entrants receive $-\kappa$, and $\OUT$ yields
$0$. Since $m\le12$, the ranking of terminal outcomes is unchanged,
so both specifications have the same pure SPNE. We use $1/m$, $-1$,
and $0$ throughout.
 
A pure strategy $\sigma_i$ assigns an action to every tree node at which player $i$ acts. A strategy profile $\sigma=(\sigma_1,\ldots,\sigma_{12})$ specifies one strategy for each player. It is an SPNE if its restriction to every subgame is a Nash equilibrium. Its equilibrium path is the ordered action sequence it induces.
For a strategy profile in $\Gamma(h)$, its \emph{outcome} is the
terminal history obtained by combining $h$ with the induced actions
of the remaining players. All equilibria considered below are in pure strategies.
At a nonterminal history $h$, an action $a$ \emph{guarantees} its mover
a payoff of at least $b$ if that player's payoff is at least $b$ in
the outcome of every SPNE of $\Gamma(h,a)$.

A \emph{singleton} is a candidate alone at its position; a \emph{copy} is an entrant choosing the position of an existing candidate. The \emph{configuration} at a history is the multiset of positions
chosen by its candidates (that is, multiplicities are taken into account). It omits action order and $\OUT$ actions.
For a history $h$ and a terminal extension of $h$, \emph{old candidates}
are the players who entered in $h$, and \emph{newcomers} are those who
enter after $h$ in that extension. The reference history is specified
in each statement or at the beginning of the relevant case. A terminal outcome \emph{retains} an old candidate if that candidate is a winner in that outcome. An all-$\OUT$ path means that every remaining player chooses $\OUT$ on the equilibrium path; their strategies still specify actions off that path.

The game described above can be easily generalized for an arbitrary number of players $n\geqslant 2$. In the section entitled ``Conjecture'', \citet{osborne} writes that for every number of players $n$, the game has ``a unique subgame perfect equilibrium outcome'', and specifies that players $1$ and $n$ enter at the median while everyone else stays out. For the present game $\Gamma$ the conjectured outcome is the action vector
\begin{equation}\label{eq:canonical}
 a^{\mathrm{can}}=(1/2,\OUT,\OUT,\OUT,\OUT,\OUT,\OUT,\OUT,\OUT,\OUT,\OUT,1/2).
\end{equation}

Our main result is the following statement.

\begin{theorem}\label{thm:main}
The game $\Gamma$ has a pure SPNE whose outcome differs from $a^{\mathrm{can}}$.
\end{theorem}

The theorem leaves open whether $a^{\mathrm{can}}$ can also be supported. %An SPNE of $\Gamma(1/2,1/10)$ with a positive payoff for player~2 does not by itself exclude $a^{\mathrm{can}}$, because a unilateral deviation holds opponents' strategies fixed. 

Spatial electoral competition builds on the location analysis of \citet{hotelling} and the theory of party competition developed by \citet{downs}. In the standard one-dimensional model with two office-motivated candidates, the candidates choose their platforms simultaneously, and voters support a nearest candidate. Both candidates locating at the median is a Nash equilibrium. Participation is fixed in this model. The conjecture studied here asks whether sequential, optional participation recovers the same two-candidate median outcome. \citet{osborne-survey} surveys spatial explanations of candidate numbers and positions under plurality rule.

Models of endogenous participation differ in the choices available to players. In the citizen-candidate frameworks of \citet{osborne-slivinski} and \citet{besley-coate}, citizens decide whether to stand and the winner implements that citizen's preferred policy. Entry therefore selects among policies tied to candidate identities. In our model, players are office-motivated and can commit to any platform if they enter. This distinction determines both the available deviations and the interpretation of entry deterrence.

Sequential location also arises in industrial organization. \citet{prescott-visscher} study firms choosing locations in sequence while anticipating later entry. \citet{loertscher-muehlheusser} analyze costly sequential entry with nonuniform consumer distributions and examine the special role of the uniform case. In their model, firms earn profits from their market shares; the present electoral model instead rewards only candidates tied for the largest vote share. This difference makes the identities and number of tied winners central to the entry incentives studied here.

The closest models also make entry sequential. \citet{palfrey} studies established parties choosing positions in anticipation of a vote-maximizing third entrant; anticipated entry can support separation of the established parties. \citet{osborne-entry} allows potential candidates to choose when to enter and commit to a position. With three potential candidates, his temporal model yields one entrant in every equilibrium under its stated assumptions. The model considered in this paper instead gives each player a single opportunity to enter, in a fixed order. These differences in timing and candidate objectives affect the responses that can deter earlier entry.

\citet{osborne} traces the conjecture to joint work with Amoz Kats in the mid-1980s. It predicts that the first and last players enter at the median and all intervening players stay out. \citet{osborne-ties} obtains a related positive result when electoral ties are restricted to candidates at the same position: under that paper's conditions, the first and last players enter at an admissible position closest to the median. Our construction uses ties among candidates at distinct positions, which the present model permits.

The closest antecedent work is \citet{devries}, who proposes a 12-player, 6-winner construction after the same first two entries.  In the final ``history'' paragraph of his online statement, \citet{osborne} reports that the proposed ``counterexample'' remained unconfirmed despite subsequent collaborative work.
We retain the first two actions proposed by \citet{devries} and provide the equilibrium verification for the subgames that follow. 

The remainder of the paper is organized as follows. Section~\ref{sec:selection} describes the plan of the proof. Sections~\ref{app:geometry}--\ref{app:solo-outer} develop the geometric arguments and establish the subgame equilibrium results used in the proof. Section~\ref{sec:positive-continuation} completes the proof, and Section~\ref{sec:conclusion} concludes.

\section{Proof strategy}\label{sec:mechanism}\label{sec:selection}
The main step is to establish the existence of an SPNE of
$\Gamma(1/2,1/10)$ in which player~2 receives a positive payoff. We first state what must be established, then describe how the
proof supplies it. Note that the action $a_3=9/10$ places players~2 and~3 symmetrically around player~1's position $1/2$.

\paragraph{The two requirements.}\label{sec:proof-requirements}
For each action $a_3\in[0,1]\cup\{\OUT\}$ of player~3 in
$\Gamma(1/2,1/10)$, we will choose an SPNE strategy profile
of $\Gamma(1/2,1/10,a_3)$.
These profiles must satisfy two requirements:
\begin{enumerate}[label=(\roman*),nosep]
\item The profile chosen for $\Gamma(1/2,1/10,9/10)$
induces an outcome in which players~2 and~3 receive
the same payoff $\rho\ge1/10$.
\item For every action $a_3$, if the outcome of the profile
chosen for $\Gamma(1/2,1/10,a_3)$ gives player~3
a payoff above $1/10$, then player~2 is also a winner
and receives the same payoff as player~3.
\end{enumerate}

To see why this suffices, fix these profiles and let player~3 choose
a best response, preferring $9/10$ in a tie. If player~3 chooses $9/10$,
player~2 receives $\rho$. Otherwise player~3's payoff is strictly above
$\rho\ge1/10$, so (ii) makes player~2 a winner with the same payoff.
Thus player~2 receives at least $1/10$ in the resulting SPNE.
Completing the profiles at earlier subgames and choosing best
responses for players~2 and~1 then gives an outcome different from
$a^{\mathrm{can}}$: player~2 enters if the first action is $1/2$, and any other
first action already differs from $a^{\mathrm{can}}$.

In the outline below, \emph{new candidates} are players who enter
following the first 2 actions $(1/2,1/10)$; this includes player~3
whenever that player enters.

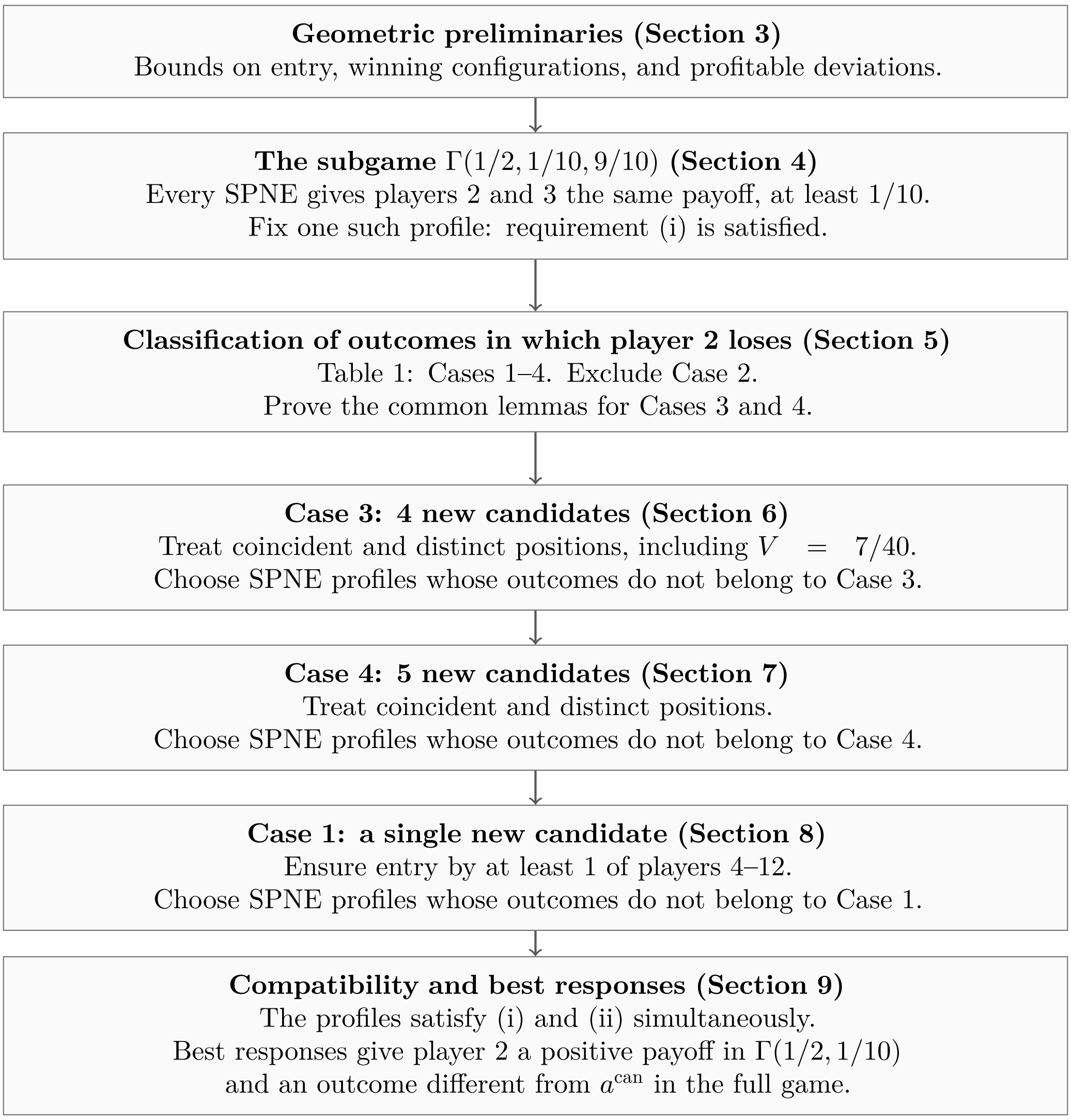
\begin{figure}[!ht]
\centering
\begin{tikzpicture}[font=\small,
 box/.style={draw=black!55,fill=black!2,align=center,
 text width=13.2cm,minimum height=1.15cm,inner sep=6pt},
 flow/.style={->,thick,draw=black!65}]
\node[box] (tools) at (0,0)
 {\textbf{Geometric preliminaries (Section~\ref{app:geometry})}\\
 Bounds on entry, winning configurations, and profitable deviations.};
\node[box] (payoffs) at (0,-1.85)
 {\textbf{The subgame $\Gamma(1/2,1/10,9/10)$ (Section~\ref{app:attainment})}\\
 Every SPNE gives players 2 and 3 the same payoff, at least $1/10$.\\
 Fix one such profile: requirement (i) is satisfied.};
\node[box] (classification) at (0,-4.1)
 {\textbf{Classification of outcomes in which player 2 loses (Section~\ref{app:strict})}\\
 Table~\ref{tab:improvement-families}: Cases 1--4. Exclude Case 2.\\
 Prove the common lemmas for Cases 3 and 4.};
\node[box] (four) at (0,-6.35)
 {\textbf{Case 3: 4 new candidates (Section~\ref{sec:case-three})}\\
 Treat coincident and distinct positions, including $V=7/40$.\\
 Choose SPNE profiles whose outcomes do not belong to Case 3.};
\node[box] (five) at (0,-8.4)
 {\textbf{Case 4: 5 new candidates (Section~\ref{sec:case-four})}\\
 Treat coincident and distinct positions.\\
 Choose SPNE profiles whose outcomes do not belong to Case 4.};
\node[box] (solo) at (0,-10.45)
 {\textbf{Case 1: a single new candidate (Section~\ref{app:solo-outer})}\\
 Ensure entry by at least 1 of players 4--12.\\
 Choose SPNE profiles whose outcomes do not belong to Case 1.};
\node[box] (result) at (0,-12.6)
 {\textbf{Compatibility and best responses (Section~\ref{sec:positive-continuation})}\\
 The profiles satisfy (i) and (ii) simultaneously.\\
 Best responses give player 2 a positive payoff in $\Gamma(1/2,1/10)$\\
 and an outcome different from $a^{\mathrm{can}}$ in the full game.};
\draw[flow] (tools)--(payoffs);
\draw[flow] (payoffs)--(classification);
\draw[flow] (classification)--(four);
\draw[flow] (four)--(five);
\draw[flow] (five)--(solo);
\draw[flow] (solo)--(result);
\end{tikzpicture}
\caption{Scheme of the proof. Case numbers refer to Table~\ref{tab:improvement-families};
the arrows show the order in which the cases are treated.}\label{fig:proof-route}
\end{figure}

Section~\ref{app:geometry} provides the geometric ideas that will be used in the proof. The following
steps explain how Sections~\ref{app:attainment}--\ref{sec:positive-continuation} establish the 2 requirements (i) and (ii); Figure~\ref{fig:proof-route} illustrates the complete proof path. Section~\ref{sec:positive-continuation} completes the argument.

%\FloatBarrier

\paragraph{1. Payoffs in $\Gamma(1/2,1/10,9/10)$.}
Write $\mathcal S=\Gamma(1/2,1/10,9/10)$; players~4--12 remain in this subgame.
Section~\ref{app:attainment} proves that every SPNE of $\mathcal S$
gives players~2 and~3 the same payoff, at least $1/10$
(Proposition~\ref{prop:attainment}). Choosing any such profile
establishes (i).

\paragraph{2. Identify the outcomes that violate (ii).}
An outcome is called \emph{harmful} if player~3 receives more than $1/10$
and player~2 loses. In a harmful outcome induced by an SPNE of
$\Gamma(1/2,1/10,a_3)$, every new candidate is a winner.
Since player~3 receives a payoff strictly above $1/10$,
there are at most 9 winners and therefore at most
9 new candidates, including player~3. Proposition~\ref{prop:improvement-classification} in
Section~\ref{sec:harmful-classification} reduces the possibilities
to 1, 4, or 5 new candidates including player 3.
Unless player~3 is the sole new candidate, the geometry requires 2
new candidates between $1/10$ and $1/2$ and at least 2 more at or
to the right of $1/2$, excluding outcomes with 2 or 3 new candidates. It also forces each
new candidate's vote share above $3/20$, while player~2 retains more than
$1/10$ of the votes, excluding outcomes with 6 or more new candidates.
Table~\ref{tab:improvement-families} lists the remaining configurations, including coincident positions.

\paragraph{3. Resolve Cases 2--4.}
A configuration with 4 or 5 new candidates is called
\emph{strict} if these candidates choose distinct positions different
from $1/10$ and $1/2$, and players 1 and 2 lose.
A useful deviation preserves a new candidate's own vote share while
removing another candidate from the winners
(Lemma~\ref{lem:deletion}). This candidate then shares victory
with fewer rivals. Entry capacities and the number of remaining
decisions determine whether later players can undo this improvement.
Section~\ref{sec:copied-families} excludes Case 2 by a final-player
deviation. The remaining parts of Section~\ref{app:strict} establish
common lemmas for Cases 3 and 4.
Section~\ref{sec:case-three} treats all of Case 3, including the
configurations in which player~1 also wins.
Section~\ref{sec:case-four} selects SPNE profiles avoiding Case 4.

\paragraph{4. Resolve Case 1: prevent player 3 from being the only new candidate.}
Call such an outcome \emph{solo}. The classification restricts
player~3's position to $[1/2,4/5]$. Section~\ref{app:triples}
excludes all-$\OUT$ paths by profitable later entry on $[1/2,7/10]$
and at $4/5$. For $7/10<a_3<4/5$,
Section~\ref{so:assembly-section} establishes the existence of an SPNE
with further entry (Corollary~\ref{so:further-entry}). These arguments provide
the profiles needed to avoid solo outcomes.

\paragraph{5. Verify compatibility and complete the equilibrium.}
Section~\ref{sec:positive-continuation} checks that all the required
choices can be made simultaneously, retaining the chosen SPNE of
$\Gamma(1/2,1/10,9/10)$ in which players~2 and~3 receive the same
payoff $\rho\ge1/10$. Table~\ref{tab:case-coverage} records which configurations are
excluded in every SPNE and which require a particular choice.
This establishes (ii). Applying the best-response argument above
gives Proposition~\ref{prop:positive-child} and Theorem~\ref{thm:main}.

\section{Geometric preliminaries}\label{app:geometry}
In this section, we collect the geometric arguments used throughout the proof.
They bound vote shares and the number of players who enter subsequently, and identify
deviations whose gains survive later entry. Section~\ref{app:attainment}
first applies these tools to $\Gamma(1/2,1/10,9/10)$.
Section~\ref{app:strict} then classifies the third player's possible
deviations in $\Gamma(1/2,1/10)$.

\paragraph{Notation for the geometric arguments.}
We write configuration multisets in braces. Histories are denoted
by $h$, $\eta$, or indexed variants; named configurations use
calligraphic letters. Position labels are lowercase, except for
the interval bounds $L,U$. Unless a particular
history is specified, a statement for a configuration and a number of
remaining players quantifies over every history of $\Gamma$ with
those features. The count includes all
remaining players, including those who subsequently choose $\OUT$.
An \emph{interior gap} is an open interval $(l,r)$ with candidates
at $l$ and $r$ and no candidates between them. If $l$ and $r$ are
the positions of the leftmost and rightmost candidates, respectively,
we call $[0,l)$ the \emph{left exterior region} and $(r,1]$ the
\emph{right exterior region}. A \emph{strict entry} is entry at a
position different from those of all existing candidates, and a
\emph{strict pair} consists of 2 new candidates at distinct positions
in the same region, each alone at its position.

At a history $h$, the \emph{entry capacity} of a position $x$ is the
vote share an additional candidate at $x$ would receive with all
existing positions fixed. This geometric quantity is defined even
at terminal histories. For a player's entry action, it is the vote
share immediately after entry. For an open region we use the supremum
of these shares over its positions.

The following formulas follow directly from the voting rule. At a
history with at least 1 candidate, entry in an interior gap $(l,r)$
has capacity $(r-l)/2$. Entry at $x<l$, where $l$ is the position of
the leftmost candidate, has capacity $(x+l)/2<l$. Entry at $x>r$,
where $r$ is the position of the rightmost candidate, has capacity
$1-(x+r)/2<1-r$.
A copy at a position whose vote cell has length $M$
and contains $j$ candidates has entry capacity $M/(j+1)$. Subsequent entry can only decrease a candidate's
vote share or the entry capacity of a fixed position.

\subsection{Restrictions on entry and winning}\label{geo:entry-restrictions}
The first bounds turn an equilibrium condition into geometric restrictions:
every player who enters in the subgame must win, and there are only
12 players.
Capacities then identify entries that cannot affect a specified incumbent.

\begin{lemma}\label{lem:allwin}
Let $h$ be any history of $\Gamma$. In the outcome of every SPNE of
$\Gamma(h)$, every player who moves in this subgame and enters is a
winner. All such players have the same vote share.
\end{lemma}
\begin{proof}
A player who entered and lost would receive $-1$. Choosing $\OUT$
instead would give that player zero, regardless of subsequent actions.
\end{proof}

\begin{lemma}\label{lem:floor}
Let $N\ge1$ be an integer. In any nonempty configuration with
at most $N$ candidates, the maximal vote share $w$ satisfies
$w\ge1/N$. If a union of vote cells in this configuration has
length $M$ and contains $k\ge1$ candidates, then $w\ge M/k$,
with equality if and only if all these candidates are winners.
\end{lemma}
\begin{proof}
The candidates' vote shares sum to $1$, giving $1\le Nw$.
The vote shares of the $k$ candidates in the stated union sum
to $M$. Each is at most $w$, so $M\le kw$, with equality
if and only if each equals $w$.
\end{proof}

\begin{lemma}\label{lem:capacity}
Let $h$ and $h'$ be histories of $\Gamma$, with $h'$ extending $h$.
The vote share of every candidate who entered in $h$ is no greater
at $h'$ than at $h$. For every $x\in[0,1]$, the entry capacity of $x$
is also no greater at $h'$ than at $h$.
\end{lemma}
\begin{proof}
Entry at a new position can only shrink an existing candidate's vote
cell, and a copy divides the same cell among more players. Choosing
$\OUT$ leaves all vote shares unchanged. Repeating this argument
along the extension proves the first claim. Apply the same argument
with an additional candidate fixed at $x$: its vote share at each
history is the entry capacity of $x$ at that history. This proves
the second claim.
\end{proof}

\begin{lemma}\label{lem:payoff-count}
Let $h$ be a nonterminal history of $\Gamma$ at which $n\ge1$ players
have entered, and let $i$ be the next player to move. Suppose player~$i$
has an action guaranteeing a \emph{payoff} of at least $1/m$, where
$m$ is a positive integer. Then, in every SPNE of $\Gamma(h)$, player~$i$
enters and wins, there are at most $m$ winners, and at most $m-1$
players who move after $i$ enter. In each such outcome, the maximal vote share is strictly
above $1/(n+m)$.
\end{lemma}
\begin{proof}
Fix any SPNE of $\Gamma(h)$. Taking the guaranteed action against
its specified subsequent strategies gives player~$i$ at least $1/m$.
Optimality therefore gives player~$i$ a payoff of at least $1/m$ in
this SPNE. That player wins, and the number of winners is at most $m$.
Every player who enters subsequently also wins by Lemma~\ref{lem:allwin}, so at
most $n+m$ players have entered. Let $w$ be the maximal vote share in this outcome. Equality in $w\ge1/(n+m)$ would
require all $n+m$ to win, contradicting the bound of $m$ winners.
\end{proof}

%The guarantee in Lemma~\ref{lem:payoff-count} concerns the prize
%shared among winners. It need not be a lower bound on vote share.
%For example, if there are exactly 2 candidates, at $1/10$ and $1/2$,
%and only player~12 remains, entry at
%$7/10$ gives vote shares $3/10,3/10,2/5$. Player~12 wins alone and
%receives payoff 1, with vote share $2/5$.

Fix a history $h$ and a class $\mathcal{H}$ of terminal histories
extending $h$. The vote share of a candidate who has entered
by $h$ is \emph{protected at $h$ with respect to $\mathcal{H}$} if, at every terminal
history in $\mathcal{H}$, that candidate has the same vote share
as at $h$. For a candidate at position $x$, an \emph{affecting entry} is a copy
at $x$, entry in an adjacent gap, or entry beyond $x$ on a side where
$x$ is extreme. These are precisely the entries that can change that
candidate's vote share. Copies at other old positions leave it unchanged.

\begin{lemma}\label{lem:protection}
Fix a history $h$ of $\Gamma$ in which player~$j$ has entered and has vote share $v$,
and let $w>0$. Suppose every entry affecting $j$, evaluated at $h$,
has capacity strictly below $w$. Then in every terminal extension of
$h$ with maximal vote share $w$ in which all players entering after
$h$ win, candidate~$j$ still has vote share $v$. Such an extension can
therefore exist only if $v\le w$.
In particular, if every such capacity is below $1/12$, candidate~$j$'s
vote share equals $v$ in the outcome of every SPNE of $\Gamma(h)$.
\end{lemma}
\begin{proof}
All candidates who entered in $h$ remain at their positions. Only an entry affecting
$j$ at $h$ can change that candidate's cell or its multiplicity.
A new candidate choosing such a location has final vote share below $w$
by Lemma~\ref{lem:capacity}, and therefore cannot win in an outcome
whose maximal vote share is $w$. Hence no such entry occurs in the
stated extensions. Candidate~$j$ keeps share $v$, which cannot exceed
the maximal share $w$. For the SPNE conclusion, Lemma~\ref{lem:allwin}
makes every player who enters after $h$ a winner, and Lemma~\ref{lem:floor}
gives $w\ge1/12$.
\end{proof}

\paragraph{Example.}
Suppose there is exactly 1 candidate at each of $2/5,1/2,3/5$. The candidate at $1/2$
has vote share $v=1/10$. Let $w$ be the maximal vote share in any
SPNE outcome of the subgame following this history. Then $w\ge1/12$.
Entry in either adjacent gap, or a copy at $1/2$, has capacity
$1/20<1/12\le w$. By Lemma~\ref{lem:protection}, that candidate
therefore keeps vote share $v$ in every SPNE outcome of this subgame,
although the 2 extreme candidates initially have larger shares.

\subsection{Geometry of winning configurations}\label{geo:winning-configurations}
The equal-share equations restrict the number and positions of new candidates
in each old gap and exterior region. The final bound in this subsection
specializes these equations to the original positions $1/10$ and $1/2$;
it is used both to bound the payoffs of players~2 and~3 in
$\Gamma(1/2,1/10,9/10)$ and to classify player~3's improvements in
$\Gamma(1/2,1/10)$.

\begin{lemma}\label{lem:gap}
Fix a history $h$ of $\Gamma$ and a terminal extension of $h$.
Let $(l,r)$ be an interior gap at $h$. In the terminal extension,
let $z_1<\cdots<z_p$, where $p\ge1$, be the distinct positions
chosen by players entering in this gap.
Suppose there are $m_j\ge1$ candidates at $z_j$
and each candidate at these new positions receives vote share $w>0$.
Writing $z_0=l$ and $z_{p+1}=r$, we have
\[
 z_{j+1}-z_{j-1}=2m_jw
 \qquad (1\le j\le p).
\]
For $p=1$, writing $m=m_1$, we have $w=(r-l)/(2m)$.
For $p=2$, writing $m=m_1$ and $n=m_2$, we have
\[
 z_1=r-2nw,\qquad z_2=l+2mw,
 \qquad 2\max\{m,n\}w<r-l<2(m+n)w.
\]
For $p\ge3$, we have $r-l\ge4w$.
\end{lemma}

\begin{proof}
The cell at $z_j$ has mass $(z_{j+1}-z_{j-1})/2$,
divided among the $m_j$ candidates at that position. This gives the formulas for $p=1$ and $p=2$.
For $p=2$, the conditions $l<z_1$ and $z_2<r$ give
$r-l>2\max\{m,n\}w$, while $z_1<z_2$ gives
$r-l<2(m+n)w$.

If $p\ge3$, the equations for $j=1$ and $j=p$ give
$z_2-l=2m_1w\ge2w$ and $r-z_{p-1}=2m_pw\ge2w$.
Since $z_2\le z_{p-1}$,
\[
 r-l=(z_2-l)+(z_{p-1}-z_2)+(r-z_{p-1})\ge4w.
\]
\end{proof}

\begin{lemma}\label{lem:short-gaps}\label{bd:gaps}
Fix a history $h$ of $\Gamma$.
Let $(l,r)$ be an interior gap at $h$, and write $G=(r-l)/2$.
Consider a terminal extension of $h$ in which at least 1 player
enters in $(l,r)$ and every such new candidate wins with
vote share $k>0$.
If $G<2k$, these new candidates are at either
1 singleton position with $k=G$, or
2 distinct singleton positions with $G/2<k<G$.
\end{lemma}

\begin{proof}
By Lemma~\ref{lem:gap}, 3 or more positions require
$G\ge2k$, contrary to the assumption. If the new candidates are at 2 distinct positions, let their multiplicities be $m,n\ge1$.
Lemma~\ref{lem:gap} gives
$\max\{m,n\}k<G<2k$, so $m=n=1$ and $G/2<k<G$. If all new candidates are at 1 position, let its multiplicity be $m\ge1$.
Lemma~\ref{lem:gap} gives $k=G/m$.
Since $G<2k$, we have $m<2$, hence $m=1$ and $k=G$.
\end{proof}

\begin{lemma}\label{lem:two-sided-repair}
Fix a history $h$ of $\Gamma$ at which player~$i$ is the only
candidate at its position and has vote share $C$.
Let $w>0$ satisfy $C/w\notin\mathbb Z$.
Suppose that, at $h$, every interior gap adjacent to this position
has length less than $4w$ and different from $2w$, and every entry
in an adjacent exterior region has capacity below $w$. Consider any terminal extension of $h$ with maximal vote share $w$
in which every player entering after $h$ wins.
Then player~$i$'s position is not copied, and at least
$\lfloor C/w\rfloor$ adjacent interior gaps each contain exactly
2 new candidates at distinct positions.
In particular, $2w<C<3w$ requires 2 such gaps.
\end{lemma}

\begin{proof}
By Lemma~\ref{lem:capacity}, a player entering at a position
whose entry capacity at $h$ is below $w$ has final vote share
below $w$ and therefore cannot win.
Thus no entry occurs in an adjacent interior gap of length
below $2w$ or in an adjacent exterior region.

Every adjacent gap receiving entry therefore has length
in $(2w,4w)$. By Lemma~\ref{lem:short-gaps}, it contains
exactly 2 new candidates at distinct positions:
the single-position case would require a gap of length $2w$.
The formulas in Lemma~\ref{lem:gap} show that each such pair
reduces the mass of the vote cell of player~$i$'s position
by exactly $w$.
Let $j$ be the number of these gaps.
The final mass of this cell is therefore $C-jw$.

If there were $m\ge2$ candidates at player~$i$'s position, at least 1
would be new and hence a winner. Since candidates at the same position
receive the same vote share, we would have
\[
 \frac{C-jw}{m}=w,
 \qquad\text{so}\qquad
 \frac{C}{w}=j+m\in\mathbb Z,
\]
contrary to the assumption.

Without copying, player~$i$ has final vote share $C-jw\le w$.
Thus $j\ge C/w-1$. Since $j$ is an integer and
$C/w\notin\mathbb Z$, this gives $j\ge\lfloor C/w\rfloor$.
\end{proof}

\begin{lemma}\label{lem:exterior-mass}
Fix a history $h$ of $\Gamma$.
Suppose the rightmost candidate at $h$ is alone at a position $x<1$.
Consider a terminal extension of $h$ with maximal vote share $w>0$
in which every player entering after $h$ wins.
Let $M$ be the sum of the final vote shares of all candidates
at or to the right of $x$, and let $j$ be the number of players
entering after $h$ at positions strictly greater than $x$.
If $M/w\notin\mathbb Z$, then position $x$ is not copied and
\[
 j=\lfloor M/w\rfloor,
 \qquad j<\frac{1-x}{w}.
\]
\end{lemma}

\begin{proof}
If a new candidate entered at $x$, it would win.
Every candidate at $x$ would then have vote share $w$.
All candidates strictly to the right of $x$ entered after $h$
and also have vote share $w$.
Thus $M/w$ would be an integer, contrary to the assumption.

Let $v$ be the final vote share of the sole candidate at $x$.
Then
\[
 M=jw+v,\qquad 0\le v\le w.
\]
Since $M/w\notin\mathbb Z$, we have $0<v<w$.
Hence $j<M/w<j+1$, giving $j=\lfloor M/w\rfloor$.

If $j=0$, the upper bound follows from $x<1$.
If $j\ge1$, let $y>x$ be the position of the nearest candidate
strictly to the right of $x$ in the terminal history.
The vote cells of all candidates strictly to the right of $x$
together have mass $1-(x+y)/2$. Therefore
\[
 jw=1-\frac{x+y}{2}<1-x,
\]
which gives the upper bound.
\end{proof}

\begin{lemma}\label{lem:two-anchor-region}
Fix a history $h$ of $\Gamma$ whose 2 leftmost candidates are
at $1/10$ and $1/2$, with no other candidate at either position.
Consider a terminal extension of $h$ with maximal vote share
$1/12\le w<1/10$ in which every player entering after $h$ wins. Then exactly 1 player enters after $h$ at a position in $[0,1/10]$,
and this position is $2w-1/10$.
The candidate already at $1/10$ has final vote share $3/10-3w$.
At least 5 players enter after $h$ at positions in $[0,1/2]$.
\end{lemma}

\begin{proof}
Let $M$ be the sum of the final vote shares of all candidates
at or to the left of $1/10$, and let $j$ be the number of
new candidates strictly to its left.
By the reflected version of Lemma~\ref{lem:exterior-mass},
whenever $M/w\notin\mathbb Z$, position $1/10$ is not copied and
\[
 j=\lfloor M/w\rfloor,
 \qquad
 j<\frac{1}{10w}\le\frac65.
\]
In that case, $j\le1$.

If no candidate enters in $(1/10,1/2)$, then $M=3/10$ and
\[
 \frac{M}{w}=\frac{3}{10w}\in(3,18/5].
\]
This ratio is not an integer, so the preceding bounds give
both $j=3$ and $j\le1$, a contradiction.

Let $z_1<\cdots<z_p$ be the distinct positions chosen by new candidates in
$(1/10,1/2)$, with final multiplicities $m_1,\ldots,m_p$.
Every candidate at these positions has vote share $w$.
If $p$ is odd, summing the equations in Lemma~\ref{lem:gap}
over odd indices gives
\[
 \sum_{\substack{1\le i\le p\\ i\text{ odd}}}m_i
 =\frac{1}{5w}\in(2,12/5],
\]
which is impossible because the sum is an integer.

Thus $p$ is even. Write
\[
 u=\sum_{\substack{1\le i\le p\\ i\text{ even}}}m_i,
 \qquad
 v=\sum_{\substack{1\le i\le p\\ i\text{ odd}}}m_i.
\]
Summing the same equations over even and odd indices gives
\[
 z_1=\frac12-2uw,
 \qquad
 z_p=\frac1{10}+2vw.
\]
Since $1/10<z_1$ and $z_p<1/2$,
\[
 1\le u,v<\frac{1}{5w}\le\frac{12}{5},
\]
so $u,v\in\{1,2\}$.

The sum of the vote shares at or to the left of $1/10$ is
\[
 M=\frac{1/10+z_1}{2}=\frac3{10}-uw.
\]
Since $3/(10w)\in(3,18/5]$ and $u$ is an integer,
$M/w$ is not an integer.
The reflected version of Lemma~\ref{lem:exterior-mass}
therefore excludes copying at $1/10$ and gives
\[
 j=\left\lfloor\frac{3}{10w}-u\right\rfloor=3-u\le1.
\]
Together with $u\in\{1,2\}$, this implies $u=2$ and $j=1$.

Let $y<1/10$ be the position of the unique new candidate
to the left of $1/10$. Its vote share is
\[
 \frac{y+1/10}{2}=w,
\]
so $y=2w-1/10$.
The candidate already at $1/10$ has final vote share
\[
 M-w=\frac3{10}-3w.
\]

There are $u+v=2+v$ new candidates in $(1/10,1/2)$.
If $v=2$, these candidates and the candidate at $y$ already
give 5 new candidates in $[0,1/2]$.
If $v=1$, they give 4, but the part of the vote cell
of position $1/2$ lying to its left has mass
\[
 \frac{1/2-z_p}{2}=\frac15-w>w.
\]
Without a copy at $1/2$, the candidate there would have
vote share greater than $w$.
Hence $1/2$ is copied, giving at least 5 new candidates
in $[0,1/2]$.
\end{proof}

\subsection{Profitable deviations and subsequent entry}\label{geo:deviations}
We first compare payoffs before and after a change of position,
holding all other actions fixed. We then give conditions under which a deviation remains profitable despite subsequent play, and conditions under which the next player enters in every SPNE of the subgame.

\begin{lemma}\label{lem:deletion}
Let $\eta$ be a terminal history of $\Gamma$ in which
candidate~$i$ is alone at position $x$.
Let $y\ne x$ be the position of another candidate, with no candidates
strictly between $x$ and $y$. Suppose that $i$ and all candidates at $y$
win with vote share $w$.
If there are candidates on the other side of $x$,
suppose that all candidates at the nearest position on that side lose. For every sufficiently small $\varepsilon>0$, replacing $i$'s action $x$ in $\eta$ by $x+\varepsilon(y-x)$, with all other actions unchanged, strictly increases $i$'s payoff.
\end{lemma}

\begin{proof}
Choose $\varepsilon>0$ sufficiently small to preserve
the order of the candidates' distinct positions.
The move decreases the vote share of every candidate at $y$. If there are candidates on both sides of $x$, player~$i$'s
vote share remains $w$, since the length of its vote cell
depends only on the positions of its 2 neighbors.
The candidates at the other adjacent position gain votes,
but their shares remain below $w$ for a sufficiently small deviation. All remaining candidates' vote shares are unchanged.
Thus player~$i$ still wins, while the candidates at $y$
no longer win and no new winners appear.
The number of winners decreases, so $i$'s payoff increases.

If player~$i$ is the leftmost or rightmost candidate,
moving toward $y$ increases player~$i$'s vote share above $w$.
The shares at $y$ decrease and all other candidates' shares
are unchanged, so $i$ becomes the sole winner.
Since there were at least 2 winners before the move,
$i$'s payoff again strictly increases.
\end{proof}

The latter lemma applies directly to deviations by player~12.
For earlier players, the subsequent actions prescribed
by the strategy profile after the deviation must also
be taken into account.

The next lemma is used in the proofs of
Propositions~\ref{sf:main} and~\ref{bd:excluded}
to exclude certain SPNE outcomes with 5 further entrants.
In both applications, 2 of these new candidates are in each of
2 interior gaps, and the remaining new candidate is in an exterior region.
The argument combines a profitable deviation by the last entrant
with a bound on the number of players remaining to move.

\begin{lemma}\label{gl:spare-deletion}
Fix a history $h$ of $\Gamma$ and a terminal extension $\eta$
in which exactly 5 players enter after $h$.
Suppose these 5 new candidates all win with vote share $q>0$
and are at distinct positions in $(0,1)$.
Suppose that 2 of these candidates enter in one interior gap at $h$,
2 enter in a different interior gap at $h$,
and the remaining candidate enters in an exterior region at $h$.
Suppose every old candidate whose position is adjacent
to one of these 5 positions in the terminal configuration
loses in $\eta$.
Suppose also that every interior gap and exterior region at $h$
containing none of the 5 new candidates has entry capacity
below $q$, evaluated at $h$. If at most 2 players remain to move after the last of these
5 entry actions, then $\eta$ is not the outcome of any SPNE
of $\Gamma(h)$.
\end{lemma}

\begin{proof}
Suppose an SPNE of $\Gamma(h)$ induces $\eta$, and let $i$
be the last of the 5 new candidates to enter.
All subsequent players choose $\OUT$ in $\eta$, so the configuration
immediately after $i$'s action is the terminal configuration in $\eta$.

If $i$ enters in an interior gap at $h$, it can move slightly
toward the other new candidate in that gap.
By Lemma~\ref{lem:deletion}, this increases its payoff
when all other actions are fixed.
Its vote share remains $q$.
If $i$ enters in the exterior region, a small move toward
its old neighbor increases its vote share above $q$.
The neighbor's share decreases and all other shares are unchanged,
so $i$ becomes the sole winner.
Thus, in either case, the deviation is profitable if all
subsequent players still choose $\OUT$.

We next show that every entry capacity at the configuration
in $\eta$ is strictly below $q$.
For a new candidate with other candidates on both sides,
its share $q$ is the sum of the capacities of its 2 adjacent gaps.
Both capacities are positive, so each is below $q$.

For the new candidate in the exterior region, consider
the right-hand case; the left-hand case follows by reflecting
positions via $t\mapsto1-t$.
Let $r$ be the position of the rightmost candidate at $h$,
and let $x>r$ be the new candidate's position.
Then
\[
 q=1-\frac{r+x}{2},
 \qquad
 \frac{x-r}{2}<q,
 \qquad
 1-x<q.
\]
The first strict inequality uses $x<1$ and bounds entry
between $r$ and $x$.
The second uses $x>r$ and bounds entry to the right of $x$.

All remaining gaps and exterior regions are unchanged from $h$
and have capacity below $q$ by assumption.
Finally, a copy at any position with $m$ candidates,
each receiving vote share $v\le q$, has capacity
\[
 \frac{mv}{m+1}<q.
\]

These bounds involve finitely many gap capacities,
exterior capacity suprema, and copying capacities.
While the order of the candidates' distinct positions is preserved,
all these quantities depend continuously on $i$'s position.
Choose the deviation sufficiently small that every entry
capacity immediately after it remains below $q$.

After the deviation, choose 1 new candidate in each of the
3 regions at $h$ containing the original 5 new candidates,
choosing player~$i$ in its region.
Player~$i$ has vote share at least $q$, and the chosen candidates
in the other 2 regions still have vote share $q$.
The candidates who entered in $h$ separate these regions and remain
at their positions after any subsequent entry.
Thus a subsequent entry can affect the vote share of at most
1 of the 3 chosen candidates.
A copy at the position of a candidate who entered in $h$ affects none of their shares.

At most 2 players remain to move.
Consequently, in every terminal continuation after the deviation,
at least 1 chosen candidate retains vote share at least $q$.
By Lemma~\ref{lem:capacity}, every player entering after
the deviation has final vote share strictly below $q$.
Such a player therefore loses.

Lemma~\ref{lem:allwin} now implies that every SPNE of the subgame
following the deviation has all remaining players choose $\OUT$.
The deviation therefore strictly increases player~$i$'s payoff,
contradicting the assumed SPNE.
\end{proof}

The next lemma propagates a positive payoff bound backwards
along a sequence of entry actions.
At each step, the next player can obtain this bound by entering.
If further entry also ensures that certain earlier candidates win,
they receive the same payoff as that player.

\begin{lemma}\label{lem:positive-chain}
Let $m\ge0$ be an integer and let $h_0,\ldots,h_m$ be histories
of $\Gamma$, with $h_{i+1}$ obtained from $h_i$ by an entry action
for every $0\le i<m$.
For each $i$, let $P_i$ be a set of players who have already
entered at $h_i$.
For $i\ge1$, suppose $P_i$ includes the player whose entry
led to $h_i$.
Suppose that, for some $v>0$:
\begin{enumerate}[label=(\roman*),nosep]
\item for each $i<m$, every SPNE of $\Gamma(h_i)$ whose outcome
includes entry by at least 1 player after $h_i$ makes
every member of $P_i$ a winner;
\item every SPNE of $\Gamma(h_m)$ gives each member of $P_m$
a payoff of at least $v$.
\end{enumerate}
Then every SPNE of $\Gamma(h_i)$ gives each member of $P_i$
a payoff of at least $v$, for all $0\le i\le m$.
For each $i<m$, the next player to move enters in every SPNE
of $\Gamma(h_i)$.
\end{lemma}

\begin{proof}
We use backward induction on $i$.
The claim for $i=m$ is assumption (ii). Suppose the claim holds for $i+1$, and fix any SPNE of $\Gamma(h_i)$.
Consider the entry action leading from $h_i$ to $h_{i+1}$.
The fixed strategy profile restricts to an SPNE
of $\Gamma(h_{i+1})$, whether or not this action is chosen
on the equilibrium path.
By the induction hypothesis, choosing this action gives
the current player at least $v$, since that player belongs
to $P_{i+1}$. Optimality therefore gives the current player a payoff
of at least $v>0$ in the fixed SPNE.
Thus this player enters and wins.
Condition (i) now implies that every member of $P_i$ also wins.
All winners receive the same payoff, so each member of $P_i$
receives at least $v$.
Since the SPNE was arbitrary, this proves the induction step.
\end{proof}

\section{The subgame \texorpdfstring{$\Gamma(1/2,1/10,9/10)$}{Gamma(1/2,1/10,9/10)}}\label{app:attainment}

\begin{figure}[!ht]
\centering
\begin{tikzpicture}[x=12cm,y=1cm,font=\small]
\fill[blue!12] (0,0.5) rectangle (1,0.95);
\foreach \b in {0,0.2,0.4,0.6,0.8,1}
 \draw[black!45] (\b,0.5)--(\b,0.95);
\foreach \x in {0.1,0.3,0.5,0.7,0.9}
 \node at (\x,0.725) {$1/5$};
\node[anchor=west] at (0,1.25) {Vote cells of the 5-position grid};
\draw (0,0)--(1,0);
\foreach \x/\lab in {0/0,0.1/{1/10},0.3/{3/10},0.5/{1/2},0.7/{7/10},0.9/{9/10},1/1}
 {\draw (\x,0.045)--(\x,-0.045);
  \node[below] at (\x,-0.1) {$\lab$};}
\foreach \x in {0.1,0.5,0.9}
 \fill (\x,0) circle (2.2pt);
\foreach \x in {0.3,0.7}
 \draw[fill=white,thick] (\x,0) circle (2.2pt);
\node[anchor=west] at (0,-0.85)
 {$\bullet$ Old candidates in $\mathcal S$};
\node[anchor=west] at (0,-1.25)
 {$\circ$ Entries at $3/10$ and $7/10$; a copy divides a position's cell equally};
\end{tikzpicture}
\caption{The configuration obtained from $\mathcal S$ after entries
at $3/10$ and $7/10$.
With 1 candidate at each position, all 5 candidates win with
vote share and payoff $1/5$.
Adding 1 copy at each position gives 10 winners, each with
vote share and payoff $1/10$.
Lemma~\ref{at:grid-retention} shows that the initial 5 candidates
win in every SPNE of the subgame following this configuration.}
\label{fig:reference}
\end{figure}
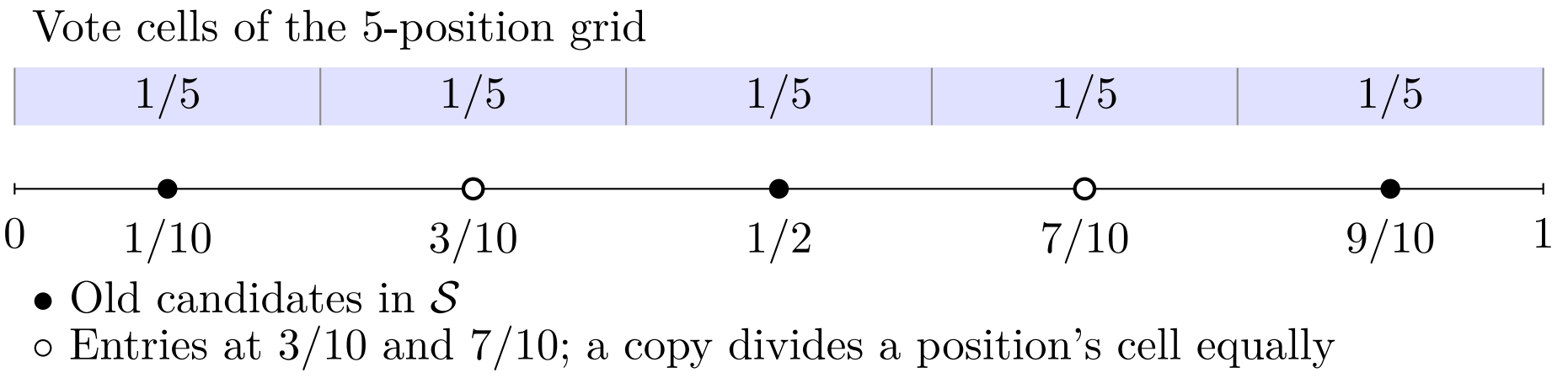

In $\mathcal S=\Gamma(1/2,1/10,9/10)$, players~4--12 remain.
We first show that player~4 has an entry action guaranteeing
a positive payoff.
We then classify the possible equilibrium outcomes and exclude
those in which players~2 and~3 lose.
The configuration obtained after additional entries at $3/10$
and $7/10$, shown in Figure~\ref{fig:reference}, provides
the initial payoff guarantee.

\begin{proposition}\label{prop:attainment}
Every SPNE of $\mathcal S=\Gamma(1/2,1/10,9/10)$ gives
players~2 and~3 the same payoff, at least $1/10$.
\end{proposition}

The remainder of this section is devoted to the proof of
Proposition~\ref{prop:attainment}. We first establish two auxiliary
lemmas that provide a positive payoff guarantee for player~4.

\Needspace{9\baselineskip}
\subsection{\texorpdfstring{Completing the grid in $\Gamma(1/2,1/10,9/10,3/10)$}{Completing the grid in Gamma(1/2,1/10,9/10,3/10)}}

\begin{lemma}\label{at:grid-retention}
Let $h$ be a history of $\Gamma$ with exactly 5 candidates,
one at each of the positions $1/10,3/10,1/2,7/10,9/10$.
In the outcome of every SPNE of $\Gamma(h)$,
each of these 5 candidates wins and receives payoff $1/5$ or $1/10$.
\end{lemma}

\begin{proof}
Fix an SPNE of $\Gamma(h)$.
If all remaining players choose $\OUT$, the 5 old candidates
each win with payoff $1/5$.

Otherwise, every player entering after $h$ wins
by Lemma~\ref{lem:allwin}.
Let $w\ge1/12$ be the maximal vote share in the outcome.
Every entry capacity at $h$ is at most $1/10$,
so Lemma~\ref{lem:capacity} gives $w\le1/10$.

Suppose $w<1/10$.
At $h$, the candidates at $3/10$ and $7/10$ each have
vote share $1/5$, and each adjacent interior gap has length $1/5$.
Moreover,
\[
 \frac{1}{5w}\in(2,12/5],
 \qquad
 w<1/10<2w.
\]
Lemma~\ref{lem:two-sided-repair} therefore requires
2 new candidates in each of the 4 gaps.
This requires 8 new candidates, but at most 7 players remain.
Hence $w=1/10$.

No player can enter in either exterior region and win,
since every such entry has capacity strictly below $1/10$ at $h$.
Let $u$ be the position of the nearest candidate strictly to the right
of $1/10$ in the terminal history.
The old candidate at $3/10$ remains there, so
\[
 1/10<u\le3/10.
\]
The cell at $1/10$ consequently has mass
\[
 \frac{1/10+u}{2}\in(1/10,1/5].
\]
Without a copy, the candidate at $1/10$ would have
vote share greater than $w$.
If there are $m\ge2$ candidates at $1/10$, the vote cell at $1/10$ must have mass
$m/10$, since every new copy wins.
The upper bound $1/5$ therefore forces $m=2$ and $u=3/10$.
Thus there is exactly 1 copy at $1/10$ and no entry in
$(1/10,3/10)$.

The cell at $3/10$ now has a left part of mass $1/10$.
Its right part has positive mass at most $1/10$,
because the old candidate at $1/2$ remains there.
The same argument forces exactly 1 copy at $3/10$
and no entry in $(3/10,1/2)$.
Applying this argument successively at $1/2$ and $7/10$
forces exactly 1 copy at each position and leaves
the remaining interior gaps empty.
Finally, the cell at $9/10$ has mass $1/5$,
so that position also has exactly 1 copy.

There are therefore exactly 10 candidates,
all with vote share $1/10$.
Each wins and receives payoff $1/10$.
\end{proof}

\begin{lemma}\label{at:grid-completion}
Let $h$ be a history of $\Gamma$ with exactly 4 candidates,
one at each of the positions $1/10,3/10,1/2,9/10$,
and 8 players remaining.
Every SPNE of $\Gamma(h)$ induces an outcome with at least
1 player entering after $h$ and gives each of the old candidates
at $1/10,3/10$ a payoff of at least $1/10$.
\end{lemma}

\begin{proof}
The next player can enter at $7/10$.
The resulting history has exactly the 5 candidates specified
in Lemma~\ref{at:grid-retention}, with 7 players remaining.
That lemma guarantees the entering player a payoff of at least $1/10$.
Thus, in every SPNE of $\Gamma(h)$, the next player enters.

Fix such an SPNE and let $w\ge1/12$ be its maximal vote share.
Every player entering after $h$ wins by Lemma~\ref{lem:allwin}.

If $w<1/10$, Lemma~\ref{lem:two-sided-repair}
applied to the candidate at $3/10$ requires
2 new candidates in each of $(1/10,3/10)$ and $(3/10,1/2)$.
These are 4 new candidates strictly to the left of $1/2$.
Reflecting positions via $t\mapsto1-t$ and applying
Lemma~\ref{lem:two-anchor-region} requires 5 more new candidates
at or to the right of $1/2$, including any copies at $1/2$.
This requires 9 new candidates, exceeding the 8 remaining players.

If $w>1/10$, every entry affecting either candidate
at $1/10$ or $3/10$ has capacity at most $1/10$ at $h$.
Lemma~\ref{lem:protection} therefore preserves both vote shares
at $1/5$, so $w\ge1/5$.
No entry capacity at $h$ exceeds $1/5$.
Since at least 1 new candidate wins, Lemma~\ref{lem:capacity}
also gives $w\le1/5$.
Thus $w=1/5$, and both old candidates win.
There are at most 5 winners, so each receives at least $1/5$.

If $w=1/10$, the argument in the proof of
Lemma~\ref{at:grid-retention} applies successively
at $1/10$ and $3/10$.
It forces exactly 1 copy at each position and preserves
both old candidates as winners.
There are at most 10 winners, since each has vote share $1/10$.
Each of the 2 old candidates therefore receives at least $1/10$.
\end{proof}

\subsection{\texorpdfstring{Payoffs of players 2 and 3 in $\mathcal S$}{Payoffs of players 2 and 3 in S}}

\begin{proof}[Proof of Proposition~\ref{prop:attainment}]
We proceed in 3 steps.

\paragraph{1. Further entry is unavoidable.}
In $\mathcal S$, player~4 can enter at $3/10$ and guarantee
a payoff of at least $1/10$ by Lemma~\ref{at:grid-completion}.
Thus player~4 enters in every SPNE of $\mathcal S$.
There are 9 players remaining after the initial history
$(1/2,1/10,9/10)$.

\paragraph{2. Classify the possible winning configurations.}
Fix an SPNE of $\mathcal S$ and let $w\ge1/12$ be
the maximal vote share in its outcome.
At least 1 player enters after the initial history,
and every such player wins by Lemma~\ref{lem:allwin}.

\smallskip
\noindent\emph{Case $w<1/10$.}
Apply Lemma~\ref{lem:two-anchor-region} to the positions
$1/10,1/2$, and then apply it after reflecting all positions
via $t\mapsto1-t$.
Players~2 and~3 each have final vote share $3/10-3w$.
There are at most 9 new candidates, each with vote share $w$,
and player~1 has vote share at most $w$.
Consequently,
\[
 1\le2(3/10-3w)+10w=3/5+4w<1,
\]
a contradiction.
This argument also covers copies of existing candidates.

\smallskip
\noindent\emph{Case $w=1/10$.}
No new candidate enters in an exterior region,
since every such entry has capacity strictly below $1/10$
at the initial history.
Without a copy at $1/10$, the candidate there would receive
the outward mass $1/10$ and a positive inward mass,
giving a vote share greater than $w$.
Thus $1/10$ must be copied.

Let $m$ be the final number of candidates at $1/10$.
Its cell has mass $m/10$ and at most $3/10$,
so $m=2$ or $m=3$.
If $m=2$, the nearest candidate strictly to the right of $1/10$ must be at $3/10$.
Let $j$ be the number of candidates at $3/10$ and let
$z\le1/2$ be the position of the nearest candidate strictly to the right of $3/10$.
Every candidate at $3/10$ is new and wins, so
\[
 z-1/10=j/5.
\]
Since $3/10<z\le1/2$, this forces $j=2$ and $z=1/2$.
If $m=3$, the nearest candidate strictly to the right of $1/10$ is at $1/2$,
so no candidate enters in $(1/10,1/2)$.

The same argument applies on the right.
The cell at $1/2$ therefore has mass $1/5$, $3/10$, or $2/5$.
Without a copy, player~1 would have vote share greater than $w$.
There must consequently be 2, 3, or 4 candidates at $1/2$,
respectively.
Every candidate then has vote share $1/10$,
giving exactly 10 winners.

\smallskip
\noindent\emph{Case $w>1/10$.}
Again, no new candidate enters in an exterior region.
Every entry capacity at the initial history is at most $1/5$,
so $w\le1/5$.
Each of the gaps $(1/10,1/2)$ and $(1/2,9/10)$
has half-length $1/5<2w$.
By Lemma~\ref{lem:short-gaps}, each gap receiving entry contains
either 1 singleton, with $w=1/5$, or
2 distinct singletons, with $w<1/5$.

Neither gap can remain empty.
To see this, suppose $(1/10,1/2)$ is empty;
the other case follows by reflection.
The cell at $1/10$ then has mass $3/10>w$,
so the position must be copied.
Since every new copy has vote share $w>1/10$,
there can be only 2 candidates at that position, giving $w=3/20$.
At this vote share, the other gap is either empty
or contains 2 distinct singletons.
The cell at $1/2$ accordingly has mass
\[
 2/5
 \qquad\text{or}\qquad
 2/5-w=1/4.
\]
Both masses exceed $w$, and neither is an integer multiple of $w$.
Thus player~1 cannot remain alone at $1/2$,
but a new copy there cannot win either.
This contradiction shows that both gaps receive entry.

If either gap contains a singleton, then $w=1/5$
and the other gap also contains a singleton.
No old position can be copied:
the extreme cells have mass at most $3/10$,
and the median cell has mass strictly below $2/5$,
so a new copy would receive less than $1/5$.
Let $x\in(1/10,1/2)$ and $y\in(1/2,9/10)$
be the 2 new positions.
The vote shares of players~2, 3, and~1, respectively,
must not exceed $1/5$, giving
\[
 x\le3/10,\qquad y\ge7/10,\qquad y-x\le2/5.
\]
Hence $x=3/10$ and $y=7/10$.
All 5 candidates win with vote share $1/5$.

Otherwise, each gap contains 2 distinct singletons.
By Lemma~\ref{lem:gap}, each pair reduces by $w$ the mass of
the vote cell at each old position bordering its gap.
The cells at $1/10$ and $9/10$ therefore have mass $3/10-w$,
and the cell at $1/2$ has mass $2/5-2w$.
If any old position were copied, each candidate there would receive
at most half its cell mass. However,
\[
 \frac{3/10-w}{2}<w,
 \qquad
 \frac{2/5-2w}{2}<w,
\]
since $w>1/10$.
Thus no old position is copied.
The vote share bound at either extreme gives
$3/10-w\le w$, hence $w\ge3/20$.

Write $t=4-20w$, so $0<t\le1$.
The 4 new positions are
\[
 (1+t)/10,\quad (5-t)/10,\quad
 (5+t)/10,\quad (9-t)/10,
 \qquad w=(4-t)/20.
\]
Players~2 and~3 each have vote share
\[
 E=(2+t)/20,
\]
and player~1 has vote share $t/10$.
At $t=1$, players~2 and~3 win together with the 4 new candidates.
For $0<t<1$, the 4 new candidates are the only winners.

\paragraph{3. Exclude the configurations in which both extremes lose.}
\label{at:four-current}\label{at:four-low-matching}
\label{at:four-current-3}\label{at:four-current-4}
\label{at:four-current-4-1}\label{at:four-current-4-2}
\label{at:four-current-5}

Suppose $0<t<1$, and let $i$ be the last of the 4 new candidates
to enter.
At most 5 players remain after its entry.
The configuration is symmetric under $x\mapsto1-x$,
so it suffices to consider a last entrant in the right pair.

If $i$ chooses the inner position $(5+t)/10$,
replace its action by $7/10$.
If it chooses the outer position $(9-t)/10$,
replace its action by $4/5$.
Both replacements preserve the order of the candidates' distinct positions
and leave the mover's vote share equal to $w$.

For the inner move, the former partner's share becomes $1/10$
and player~1's share becomes $E$.
For the outer move, the former partner's share becomes $3/20$
and player~3's share becomes $3/20$.
All these shares are strictly below $w$ because $0<t<1$.
All other candidates' shares are unchanged.
Thus each replacement leaves exactly 3 winners.

Let $h'$ be the history immediately after the replacement.
We show that every SPNE of $\Gamma(h')$ has all remaining
players choose $\OUT$.
Suppose instead that a terminal extension of $h'$ has at least
1 subsequent entrant and every such entrant wins.
Let $k\ge1/12$ be its maximal vote share.
All capacities in the following argument are evaluated at $h'$.

\smallskip
\noindent\emph{Case $k\ge1/10$.}
Every entry affecting the candidate at $9/10$ has capacity
strictly below $1/10$.
Its inward and copying capacities are below $1/10$,
and every entry to its right has capacity strictly below $1/10$.
By Lemma~\ref{lem:protection}, this candidate keeps
vote share $E$ after the inner move and $3/20$ after the outer move.
Hence $k\ge E$ in the first case and $k\ge3/20$ in the second.

The capacities of entries affecting the mover are at most
$1/10$ after the inner move and $(3-t)/20$ after the outer move.
These bounds are strictly below $E$ and $3/20$, respectively.
Lemma~\ref{lem:protection} therefore preserves
the mover's vote share $w$, so $k\ge w$.

On the other hand, every entry capacity at $h'$ is strictly below $w$.
Indeed, the interior gap capacities are at most
\[
 \frac{2-t}{10}<w,
\]
the exterior capacity suprema are $1/10<w$,
and copying capacities are at most $w/2<w$.
Lemma~\ref{lem:capacity} implies that every subsequent entrant
has final vote share below $w\le k$ and therefore loses.
This contradicts the assumed winning of all subsequent entrants.

\smallskip
\noindent\emph{Case $1/12\le k<1/10$.}
Consider the following 4 groups of candidates present at $h'$,
listed from left to right:
\begin{enumerate}[label=(\roman*),nosep]
\item the candidate at $1/10$;
\item the 2 candidates in the left pair;
\item after the inner move, the candidates at
$1/2,7/10,(9-t)/10$; after the outer move,
the candidates at $(5+t)/10,4/5$;
\item the candidate at $9/10$.
\end{enumerate}
For each group, take the region formed by the union of its
vote cells at $h'$.
After the outer move, the cell of the candidate at $1/2$
is not included in these regions.

The gaps separating these regions have capacities
$t/20$ or $1/20$, both below $1/12$.
No subsequent winner can enter in such a gap.
The boundaries of the regions therefore remain fixed,
and each region continues to be a union of whole vote cells.

Their masses and numbers of candidates already present at $h'$
are as follows.
Each table entry is a pair consisting of the mass
and the number of these candidates.
\[
\begin{array}{c|cc}
 & \text{Move to }7/10 & \text{Move to }4/5\\
\hline
\text{Group (i)}   & (E,1)   & (E,1)\\
\text{Group (ii)}  & (2w,2)  & (2w,2)\\
\text{Group (iii)} & (2/5,3) & ((7-t)/20,2)\\
\text{Group (iv)}  & (E,1)   & (3/20,1)
\end{array}
\]
Since
\[
 E>1/10,\qquad 2w>3/10,\qquad
 (7-t)/20>3/10,\qquad k<1/10,
\]
Lemma~\ref{lem:floor} requires at least 1, 2, 2, and 1
additional candidates in these 4 disjoint regions, respectively.
For the inner move, the third region has mass $2/5$
and requires at least 5 candidates in total,
of whom only 3 are already present.
Thus both moves require at least 6 subsequent entrants,
exceeding the 5 remaining players.

We have excluded every terminal extension of $h'$
in which at least 1 subsequent player enters and all such players win.
By Lemma~\ref{lem:allwin}, every SPNE of $\Gamma(h')$
therefore has all remaining players choose $\OUT$.
The deviation gives player~$i$ payoff $1/3$ instead of $1/4$,
contradicting the assumed SPNE.
This excludes all configurations with $0<t<1$.

The remaining cases have maximal vote share $1/5$, $3/20$,
or $1/10$, with 5, 6, or 10 winners, respectively.
Players~2 and~3 win in every case and receive the same payoff
\[
 \rho\in\{1/5,1/6,1/10\}.
\]
These are necessary payoff values; the argument does not assert
that each value is attained by an SPNE.
In particular, $\rho\ge1/10$.

Existence follows from the finite-payoff backward-induction
argument in Section~\ref{sec:positive-continuation}.
\end{proof}

We can now fix an SPNE of $\mathcal S$ and denote the payoff
of each of players~2 and~3 by $\rho\ge1/10$.
This establishes requirement (i) in Section~\ref{sec:mechanism}.
The next sections treat the other actions of player~3
in order to satisfy requirement (ii).

\section{Classification of outcomes in which player 2 loses}\label{app:strict}
After $(1/2,1/10)$, we classify SPNE outcomes in which player~3
receives more than $1/10$ and player~2 loses.
Table~\ref{tab:improvement-families} divides them into Cases 1--4.
This section excludes Case 2 and proves the common lemmas needed
for Cases 3 and 4. The subsequent sections treat each remaining
case in full before the proof of the main result is completed.

\subsection{\texorpdfstring{$\Gamma(1/2,1/10)$: possible outcomes when player 2 loses}{Gamma(1/2,1/10): possible outcomes when player 2 loses}}\label{sec:harmful-classification}
Throughout this subsection, new candidates are defined relative to the
history $(1/2,1/10)$: they are players~3--12 who choose positions.
In particular, player~3 is counted whenever $a_3\ne\OUT$.
Player~3's action is fixed; only the strategies of players~4--12 are
required to form an SPNE of the resulting subgame.

For the description in Table~\ref{tab:improvement-families}, let $V$
denote the maximal vote share in the terminal outcome and put
\[
 a=1/2-2V,\qquad b=1/10+2V,\qquad s=1/2+2V.
\]
The case numbers identify the rows of Table~\ref{tab:improvement-families};
the count is the number of new candidates, including player~3.
For Cases 3 and 4, $r$ denotes the leftmost new position to the
right of $1/2$; its formula depends on the number of new candidates.
Repeated positions in a multiset represent different candidates at the
same position. Each payoff in the table is the payoff of a new candidate.
A configuration with 4 or 5 new candidates is called \emph{strict}
if their positions are distinct and different from $1/10$ and $1/2$,
and players~1 and~2 lose.

\begin{proposition}\label{prop:improvement-classification}
Let $a_3\in[0,1]$ and fix an SPNE of $\Gamma(1/2,1/10,a_3)$.
If its outcome gives player~3 a payoff greater than $1/10$ and
player~2 loses, then there are 1, 4, or 5 new candidates.
Their positions and payoffs satisfy the corresponding restrictions
in Table~\ref{tab:improvement-families}.
\end{proposition}

\begingroup
\normalfont
\begin{longtable}{@{}>{\raggedright\arraybackslash}p{0.06\textwidth}>{\raggedright\arraybackslash}p{0.08\textwidth}>{\raggedright\arraybackslash}p{0.46\textwidth}>{\raggedright\arraybackslash}p{\dimexpr0.40\textwidth-6\tabcolsep\relax}@{}}
\caption{Possible positions of new candidates when player~3 receives a payoff greater than $1/10$ and player~2 loses.}\label{tab:improvement-families}\\
\toprule \textbf{Case} & \textbf{Count} & \textbf{New candidates' positions} & \textbf{Range and winners}\\\midrule\endfirsthead
\toprule \textbf{Case} & \textbf{Count} & \textbf{New candidates' positions} & \textbf{Range and winners}\\\midrule\endhead
1 & 1 & $\{a_3\}$ & $1/2\leq a_3\leq4/5$. Payoff $1/2$ at $a_3=1/2,4/5$, and $1$ otherwise; $a_3=1/2$ is a copy.\\
2 & 4 & $\{3/20,9/20,1/2,1/2\}$ & Vote share $7/40$; 5 winners, payoff $1/5$.\\
3 & 4 & $\{a,b,r,s\}$ or $\{a,b,r,r\}$,\newline where $r=3/2-4V$ & $7/40\leq V<1/5$. At $V=7/40$ player~1 also wins (payoff $1/5$); otherwise 4 winners (payoff $1/4$).\\
4 & 5 & $\{a,b,r,s,t\}$, $\{a,b,r,s,s\}$, or $\{a,b,r,r,r\}$, where $r=3/2-6V$, $t=3/2-4V$ & $3/20<V<1/6$. 5 winners, payoff $1/5$.\\\bottomrule
\end{longtable}
\endgroup

\begin{proof}
We first show that every new candidate wins and has vote share
$V>1/10$. The equal-share equations then restrict the number and
positions of new candidates between $1/10$ and $1/2$ and to the
right of $1/2$. Finally, we consider the cases with 0, 1, or 2 new
candidates in $(1/10,1/2)$. Summing the vote shares of all candidates
at or to the right of $1/2$ excludes the case with 1 and gives the
configurations
in Table~\ref{tab:improvement-families} for the other 2 cases.

\par\medskip
\noindent\textbf{The winning vote share and the left gap.} 

Every new candidate wins: player~3 has a positive payoff by assumption,
and players~4--12 who enter win by Lemma~\ref{lem:allwin}.
If any of these candidates chose $1/10$, player~2 would win
with the same payoff. This contradicts the hypothesis that player~2
loses. Henceforth player~2 remains a singleton and loses.

Let $V$ be the maximal vote share in the terminal outcome.
Player~3's payoff exceeds $1/10$, so there are at most 9 winners.
Let $n$ be the number of new candidates, so $1\le n\le9$.
There are 2 old candidates, hence at most 11 candidates in total,
and $V\ge1/11$ by Lemma~\ref{lem:floor}.
If $V<1/10$, Lemma~\ref{lem:two-anchor-region} gives player~2
vote share $3/10-3V$. Player~1 has vote share at most $V$, and all $n$
new candidates have vote share $V$. Summing all vote shares yields
\[
 1\le 3/10-3V+(n+1)V\le3/10+7V<1,
\]
a contradiction. At $V=1/10$, no new candidate strictly to the left of $1/10$ can attain
$V$; since the position $1/10$ is not copied, its vote cell contains $[0,1/10]$ and an interval of positive length
to the right of $1/10$, giving player~2 vote share strictly above $V$, a contradiction.
Thus $V>1/10$.

For $V>1/10$, no player entering strictly to the left of $1/10$ can win. The gap $(1/10,1/2)$, of
half-length $1/5<2V$, contains 0, 1 or 2 distinct
candidates, each alone at its position, by Lemma~\ref{lem:short-gaps}.
If there is 1, then $V=1/5$; if there are 2, their positions are $a=1/2-2V,b=1/10+2V$, with $V<1/5$.

With 2 new candidates in the interior gap, player~2 has $3/10-V$, so its
loss requires $V>3/20$. With none it has $3/10$, requiring
$V>3/10$; 1 new candidate in the gap has $V=1/5$.
In particular $V\ge3/20$ throughout.

If $j$ new candidates are strictly to the right of $1/2$, their total
vote share $jV$ is below $1/2$, so $j\le3$. If $j\ge1$ and $z$ is their leftmost position, their vote cells
together form the interval $[(1/2+z)/2,1]$. Thus
\[
 jV=1-(1/2+z)/2,\qquad z=3/2-2jV.
\]
For $j=1$, the constraint $z\le1$ requires $V\ge1/4$. For $j=2$, the
candidates coincide or the second position is $1/2+2V$; the latter
requires $V>1/6$. For $j=3$, the vote-share equations in Lemma~\ref{lem:gap} give
exactly the 3 forms in Case 4: multiplicities
$(1,1,1)$, $(1,2)$ or $(3)$, all requiring $V<1/6$.
The remaining pattern $(2,1)$ ends at $1/2+4V>1$ and is impossible.
These equations include a single new candidate at $1$. Copies at $1/2$
are separate and do not change this regional calculation.

\par\medskip
\noindent\textbf{The region at and right of $1/2$.}
Let $x$ be the position of the nearest candidate strictly to the left of $1/2$ ($x=1/10$
when $(1/10,1/2)$ is empty), and let $m\ge1$ be the total multiplicity
at $1/2$. The union of the vote cells of candidates at or to the right
of $1/2$ is $[(x+1/2)/2,1]$, of length $M=3/4-x/2$. Thus
\[
 m=1:\quad M\le(j+1)V;\qquad
 m\ge2:\quad M=(m+j)V,
\]
since a new copy makes every candidate in the region a winner.

With no new candidate in $(1/10,1/2)$, $V>3/10$ and $j\le1$.
Here $M=7/10$. A copy at $1/2$ forces $m+j=2$, hence
$m=2,j=0,V=7/20$. Without a copy, the sole new candidate is
strictly right of $1/2$; its location $a_3=3/2-2V$ satisfies
$1/2<a_3\le4/5$ by $M\le2V$. These are precisely the possibilities in Case 1 of the table.

With 1 new candidate at a position $x$ in the gap, $V=1/5$ and $x<3/10$, so
$M\in(3/5,7/10)$ and $j\le2$. An uncopied candidate at $1/2$ would give
$M\le3/5$, while a copy would make $M$ an integer multiple of
$1/5$. Both are impossible.

With 2 new candidates at positions $a,b$ in the gap, $3/20<V<1/5$ and
$M=7/10-V$. If the position $1/2$ is copied, $M/V\in(5/2,11/3)$ forces
$m+j=3$ and $V=7/40$. A single new candidate strictly to the right
of $1/2$ would require $V\ge1/4$. Thus only $m=3,j=0$ remains,
giving Case 2, with 2 new candidates at $1/2$.
Without copies, $j=0,1$ are impossible; $j=2$ requires
$V\ge7/40$, with player~1 also winning exactly when $V=7/40$.
For $j=3$, the position constraints derived above give $V<1/6$,
and player~1 strictly loses. The initial strict bound $V>3/20$
excludes a tie with player~2. The right-position forms derived above
therefore give exactly Cases 3 and 4, including all
multiplicities.

\end{proof}

Section~\ref{sec:copied-families} excludes Case 2.
Sections~\ref{sec:strict-families} and~\ref{sf:selected-completion}
establish the arguments shared by Cases 3 and 4.
Section~\ref{sec:case-three} then treats all of Case 3, including
coincident positions and the case $V=7/40$ where player~1 also wins.
Section~\ref{sec:case-four} treats Case 4, and
Section~\ref{app:solo-outer} treats Case 1.
Finally, Section~\ref{sec:positive-continuation} verifies that the
required choices of SPNE profiles can be made simultaneously.

\subsection{\texorpdfstring{Case 2: 2 new candidates at $1/2$}{Case 2: 2 new candidates at 1/2}}\label{sec:copied-families}
Case 2 has 4 new candidates, at $3/20$, $9/20$, and twice at $1/2$.
We exclude it by changing player~12's action while keeping the
first 11 actions fixed.

No SPNE of $\Gamma(1/2,1/10,a_3)$ has an outcome in which the positions of the new candidates form the multiset \(\{3/20,9/20,1/2,1/2\}\). Player~12 can instead choose \(3/4\) and win alone. To evaluate this
deviation, remove player~12's candidate if that player originally entered;
if its original action was $\OUT$, remove no candidate. Each remaining
candidate then has vote share at most \(3/10\), whereas entry at \(3/4\)
gives player~12 vote share \(3/8\). This entry cannot increase any other
candidate's vote share.

\subsection{Cases 3 and 4: common deviation arguments}\label{sec:strict-families}
\begin{figure}[!ht]
\centering
\begin{tikzpicture}[x=12.8cm,y=1cm,font=\small]
  % Strict four: V=3/16 is used ONLY for horizontal spacing.
  \node[anchor=west] at (0,0.72)
    {\textbf{Case 3: 4 new candidates,} $7/40<V<1/5$};
  \fill[black!8] (0,-0.92) rectangle (.1125,-0.50);
  \fill[blue!12] (.1125,-0.92) rectangle (.4875,-0.50);
  \fill[black!8] (.4875,-0.92) rectangle (.625,-0.50);
  \fill[blue!12] (.625,-0.92) rectangle (1,-0.50);
  \draw[black!55] (0,-0.92) rectangle (1,-0.50);
  \foreach \z in {.1125,.3,.4875,.625,.8125}
    \draw[densely dashed,black!45] (\z,-0.36)--(\z,-0.92);
  \foreach \z/\lab in {
    .05625/\alpha,.20625/V,.39375/V,.55625/{\beta_4},
    .71875/V,.90625/V}
    \node[font=\scriptsize] at (\z,-0.71) {$\lab$};
  \draw (0,0)--(1,0);
  \foreach \z/\lab in {.1/{1/10},.5/{1/2}}
    {\fill (\z,0) circle (2pt);
     \node[above] at (\z,.08) {$\lab$};}
  \foreach \z/\lab in {.125/a,.475/b,.75/r,.875/s}
    {\draw[blue!70!black,thick,fill=white] (\z,0) circle (2pt);
     \node[below] at (\z,-.08) {$\lab$};}
  \foreach \z/\lab in {
    0/0,.1125/\alpha,.3/{3/10},.4875/{3/10+V},
    .625/{1-2V},.8125/{1-V},1/1}
    \node[below,font=\scriptsize] at (\z,-.98) {$\lab$};

  % Strict five: V=4/25 is used ONLY for horizontal spacing.
  \begin{scope}[yshift=-3.05cm]
    \node[anchor=west] at (0,0.72)
      {\textbf{Case 4: 5 new candidates,} $3/20<V<1/6$};
    \fill[black!8] (0,-0.92) rectangle (.14,-0.50);
    \fill[blue!12] (.14,-0.92) rectangle (.46,-0.50);
    \fill[black!8] (.46,-0.92) rectangle (.52,-0.50);
    \fill[blue!12] (.52,-0.92) rectangle (1,-0.50);
    \draw[black!55] (0,-0.92) rectangle (1,-0.50);
    \foreach \z in {.14,.3,.46,.52,.68,.84}
      \draw[densely dashed,black!45] (\z,-0.36)--(\z,-0.92);
    \foreach \z/\lab in {
      .07/\alpha,.22/V,.38/V,.49/{\beta_5},.60/V,.76/V,.92/V}
      \node[font=\scriptsize] at (\z,-0.71) {$\lab$};
    \draw (0,0)--(1,0);
    \foreach \z/\lab in {.1/{1/10},.5/{1/2}}
      {\fill (\z,0) circle (2pt);
       \node[above] at (\z,.08) {$\lab$};}
    \foreach \z/\lab in {.18/a,.42/b,.54/r,.82/s,.86/t}
      {\draw[blue!70!black,thick,fill=white] (\z,0) circle (2pt);
       \node[below] at (\z,-.08) {$\lab$};}
    \foreach \z/\lab in {
      0/0,.14/\alpha,.3/{3/10},.46/{3/10+V},
      .68/{1-2V},.84/{1-V},1/1}
      \node[below,font=\scriptsize] at (\z,-.98) {$\lab$};
    \draw[black!45] (.52,-.92)--(.52,-1.35);
    \node[below,font=\scriptsize] at (.52,-1.35) {$1-3V$};
  \end{scope}
  \fill (.015,-5.05) circle (2pt);
  \node[anchor=west,font=\scriptsize] at (.03,-5.05) {old candidate};
  \draw[blue!70!black,thick,fill=white] (.27,-5.05) circle (2pt);
  \node[anchor=west,font=\scriptsize] at (.285,-5.05) {new candidate};
  \fill[blue!12] (.51,-5.14) rectangle (.55,-4.96);
  \node[anchor=west,font=\scriptsize] at (.565,-5.05) {winning cell};
  \fill[black!8] (.79,-5.14) rectangle (.83,-4.96);
  \node[anchor=west,font=\scriptsize] at (.845,-5.05) {losing cell};
\end{tikzpicture}
\caption{The distinct-position configurations in Cases 3 and 4 with both original candidates losing.
Dashed lines mark vote-cell boundaries, and the bands show cell
masses. In both panels, player~2 has vote share $\alpha=3/10-V$; player~1
has $\beta_4=7/10-3V$ or $\beta_5=7/10-4V$, respectively.
All new candidates have vote share $V$. The horizontal spacing uses
$V=3/16$ and $V=4/25$; the position order and formulas hold throughout
the displayed ranges. The figures describe the configurations in Proposition~\ref{prop:improvement-classification};
their compatibility with SPNE is examined below.}
\label{fig:strict-families}
\end{figure}
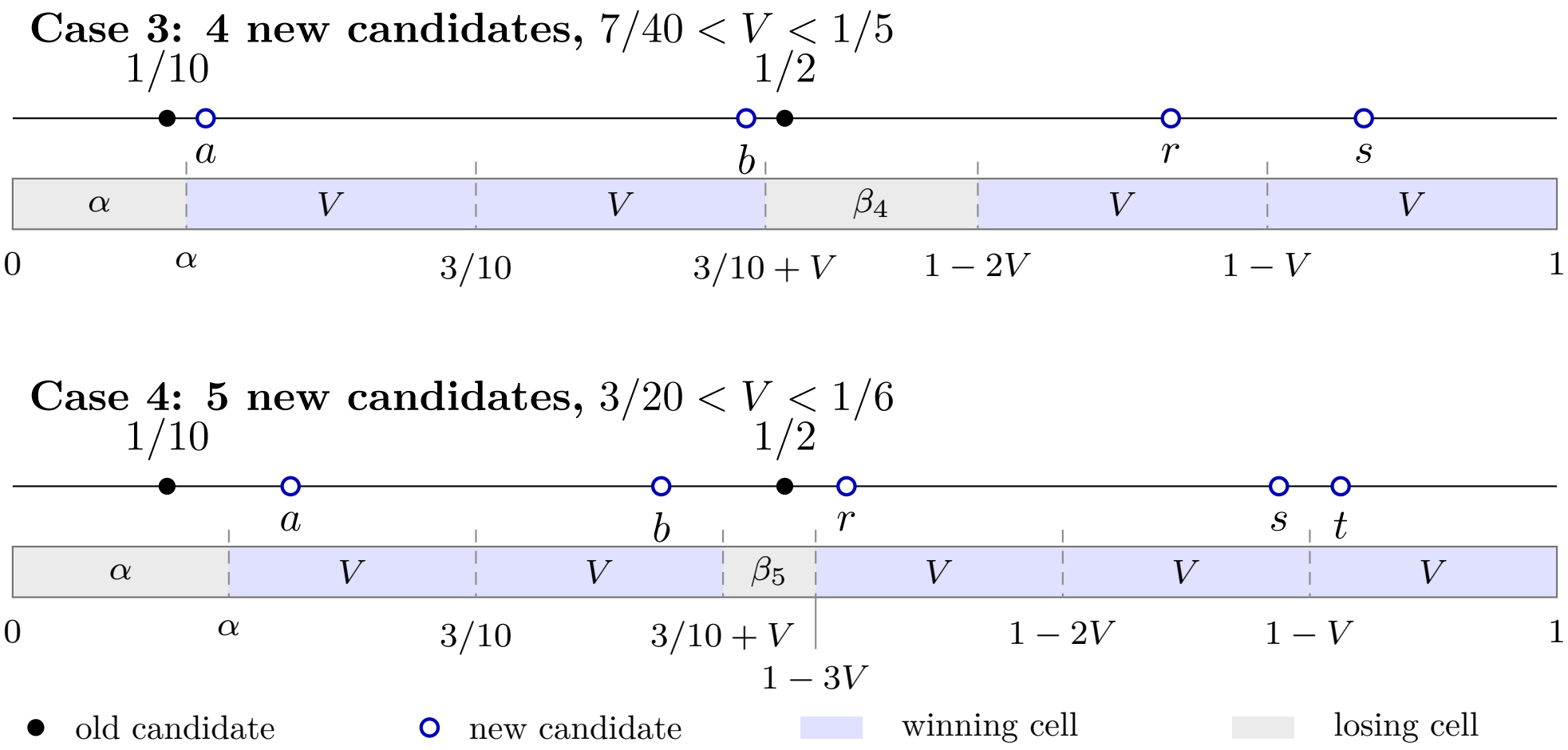

Cases 3 and 4 share 2 useful deviations. The first excludes a
copied rightmost position when the vote mass to its right is
sufficiently large. The second rules out either left position as
the first new entry in a strict configuration.
Figure~\ref{fig:strict-families} displays the distinct-position
configurations. The order of positions does not specify the order of entry.

\begin{lemma}\label{lem:copied-extreme-response}
Let $(h,a_{12})$ be a terminal history with maximal vote share $V$.
Suppose its rightmost position $z<1$ has $m\ge2$ candidates,
each with vote share $V$, and $1-z>V$.
If $a_{12}\in\{\OUT,z\}$, player~12 has a strictly profitable
deviation that makes it the sole winner.
\end{lemma}
\begin{proof}
Put $F=1-z$ and choose $\varepsilon>0$ sufficiently small that
$z+\varepsilon\le1$ and $F-\varepsilon/2>V$.
Replacing $a_{12}$ by $z+\varepsilon$ gives player~12 vote share
$F-\varepsilon/2$. The vote cell at $z$ then has length
$mV-F+\varepsilon/2<(m-1)V$.
This is divided among $m$ candidates if $a_{12}=\OUT$ and among
$m-1$ if $a_{12}=z$. Each receives less than $V$, while all other
candidates retain vote shares at most $V$.
Player~12 therefore wins alone, improving on its original payoff,
which was zero or at most $1/m$.
\end{proof}

\paragraph{Excluding a first new candidate on the left.}
The next lemma applies when exactly 3 candidates have entered:
players~1 and~2 at $1/2$ and $1/10$, and the first new candidate
at a position $c\in(1/10,1/2)$ satisfying the lemma's bounds.
For either left position $a$ or $b$ in a strict configuration,
the next player can guarantee payoff 1. The candidate at $c$
would therefore lose, contradicting the proposed outcome.
The lemma also applies at $V=7/40$ in Section~\ref{app:boundaries}, where
player~1 also wins in the configuration being examined.

\begin{lemma}\label{lem:left-interval-response}
Let $h$ be a nonterminal history of $\Gamma$ with exactly 3 candidates,
one at each of the positions $1/10,c,1/2$, where
$c\in(1/10,1/5]\cup[2/5,1/2)$. Let $i$ be the next player to move,
and suppose at most 8 players move after $i$. Define
\[
 u=\min\{c-1/10,1/2-c\},\qquad z=7/10+u/2.
\]
Every SPNE of $\Gamma(h,z)$ induces an outcome in which
player~$i$ is the sole winner and receives payoff 1.
\end{lemma}
\begin{proof}
Immediately after player~$i$ enters at $z$, that player is the sole winner. Put $G=\max\{c-1/10,1/2-c\}/2$,
$D=(z-1/2)/2$, $F=1-z$, and $R=1-(1/2+z)/2$.
The shorter gap on the left has entry capacity below $1/12$, and
\[
 1/4\le F<3/10,\quad F<2G<R,\quad R\le3F/2,\quad G/2<D<1/6,
\]
with $R=3F/2$ only when $u=1/10$.
Consider a terminal extension of $(h,z)$ in which at least 1
player moving after $i$ enters and every such player wins.
Let $k\ge1/12$ denote the maximal vote share in this extension.
In this proof, old and new candidates are defined relative to $(h,z)$.
We rule out such an extension separately for $k<1/10$, $k=1/10$,
and $k>1/10$.

For $k<1/10$, suppose first that $k>G/2$. Then $k<G<2k$,
while the old vote share $1/5$ at $c$ satisfies
$2<(1/5)/k\le12/5$. Lemma~\ref{lem:two-sided-repair}
would require 2 adjacent gaps with half-lengths in $(k,2k)$, although only 1 exists.
Consequently
\[
 k\le G/2=1/10-u/4,\qquad
 1/12\le1/10-u/4,\qquad u\le1/15.
\]
The gap $(1/2,z)$ contains either no new candidate or 2 new candidates
at distinct positions; let $\varepsilon=0$ or $1$, respectively.
The vote cells of the candidates at or to the right of $z$ have
total length $R-\varepsilon k$, with
$4<R/k<5$ and $F<4k$. If $\varepsilon=0$,
Lemma~\ref{lem:exterior-mass} requires at least 4 new candidates
in the right exterior region, whereas its length bound permits fewer than 4. Thus $\varepsilon=1$;
the same lemma then requires exactly 3 new candidates in the exterior region.
Together with the pair these are 5 new candidates strictly right of $1/2$.

For $c\le1/5$, the candidate at $1/10$ has initial vote share $C=1/10+u/2$;
for $c\ge2/5$, the candidate at $1/2$ has $C=1/10+3u/4$.
Their copy capacities satisfy $C/2\le3/40<1/12$.
No player can enter in the shorter gap on the left and win. For $c\le1/5$, $C>k$
forces exactly 1 new candidate in the left exterior region, since 2 vote shares of $k$
exceed the mass $1/10$ of voters in $[0,1/10]$. For $c\ge2/5$ the pair in $(1/2,z)$ reduces
the cell at $1/2$. In either case the indicated old vote share becomes $C-k$.

The leftmost candidate in $(1/2,z)$ is at $z-2k$, so the union
of vote cells of candidates at or to the left of $1/2$ has length
$(1/2+z-2k)/2=1-R-k$. If at most 3 new candidates lie there,
it contains the indicated old candidate and at most 5 others,
each of vote share at most $k$. Hence
\[
 1-R-k\le C-k+5k,\qquad
 1/2-u/2\le1-R-C\le5k\le1/2-5u/4,
\]
which contradicts $u>0$. At least 4 new candidates at or to the left of $1/2$ are therefore
necessary. Together with the 5 on the right, this requires 9 new
candidates, but at most 8 players remain.

At $k=1/10$ and $c\le1/5$, every entry affecting the candidate at
$1/10$ has capacity below $k$, so that candidate retains vote share
$1/10+u/2>k$, a contradiction. For $c\ge2/5$, the candidate at
$1/10$ has vote share $F$, with $2k<F<3k$; each entry to its left
has capacity strictly below $k$. Lemma~\ref{lem:two-sided-repair}
would require 2 adjacent interior gaps, whereas this candidate has only 1.

For $k>1/10$, the longer gap on the left contains no new candidate,
1 new candidate with $k=G$, or 2 new candidates at distinct positions.
If there is no entry in this gap, the candidate at $c$ retains vote
share $1/5$, so $k\ge1/5$. The candidate at $1/2$ when $c\le1/5$,
or at $1/10$ when $c\ge2/5$, then retains vote share at least $F$,
which exceeds every attainable entry capacity. This is incompatible with a winning new candidate.
For $c\le1/5$, the candidate at $1/10$ retains vote share
$(1/10+c)/2>D$. Hence $k>D$ and no new candidate can win in $(1/2,z)$.
With 2 new candidates in the longer left gap, the vote cell at
$1/2$ (if $c\le1/5$) or at $1/10$ (if $c\ge2/5$) has length
in $[F-k,3/10-k)$. A copy there would
receive less than $k$, so no such copy occurs, and $k\ge F/2$.
With 1 new candidate in this gap, $k=G>F/2$.
Except at $c=2/5,k=1/8$, therefore, $k>D$, $(1/2,z)$ is empty,
and $F\le2k<R<3k$, contradicting Lemma~\ref{lem:exterior-mass}.
At the exception the left pair leaves the candidate at $1/2$ with vote share $7/40>k$ and
copy capacity below $k$, forcing a singleton $x$ in $(1/2,z)$, whose length is $2k$.
The vote cells of the candidates at or to the right of $z$ then
have total length $1-(x+z)/2\in(2k,3k)$, while $F=2k$,
again violating that lemma. Thus all players moving after $i$ choose $\OUT$ in every SPNE of $\Gamma(h,z)$.
\end{proof}

\subsection{Cases 3 and 4: payoffs when the left pair is completed}\label{sf:selected-completion}
The next lemma gives the same payoff to the next player and the
candidates already on the right, provided that the next player
can guarantee a specified payoff. Each case will establish this
guarantee using its own deviation.
Use $a=1/2-2V$, $b=1/10+2V$ and $s=1/2+2V$, as in
Table~\ref{tab:improvement-families}.
For Case 3, let $7/40<V<1/5$, $r=3/2-4V$ and
$\mathcal R=\{r,s\}$.
For Case 4, let $3/20<V<1/6$, $r=3/2-6V$,
$t=3/2-4V$ and $\mathcal R=\{r,s,t\}$.
Put $m=1+|\mathcal R|$.

\begin{lemma}\label{lem:left-pair-completion}
Let $\eta$ extend $(1/2,1/10)$ and have exactly 1 candidate at each position in
$\{1/10,c,1/2\}\cup\mathcal R$ and no other candidates,
where $c\in\{a,b\}$, and suppose 1 to 6 players remain.
If the next player has an action guaranteeing payoff at least $1/m$,
then every SPNE of $\Gamma(\eta)$ gives the next player and each
candidate already at a position in $\mathcal R$ payoff $1/m$.
\end{lemma}
\begin{proof}
Old and new candidates in this proof are defined relative to $\eta$.
By Lemma~\ref{lem:payoff-count}, the next player enters in every SPNE,
there are at most $m$ winners, and at most $m$ players enter after $\eta$,
including the next player. Let $w$ be the maximal vote share in any
such SPNE outcome. There are $m+2$ old candidates, so
$w>1/(2m+2)\ge1/10$, since $m\in\{3,4\}$.
Every entry capacity at $\eta$ is at most $V$.
Thus $w\le V$: each new candidate's final vote share is at most
its entry capacity at $\eta$.
For $w<V$, reducing the vote share of the old candidate at $c$
to at most $w$ requires 2 new candidates in its long gap:
$V<2w$ and its copy capacity is $1/10<w$.
Reducing the vote share of the rightmost candidate at a position in
$\mathcal R$ to at most $w$ requires entry strictly to its right, giving
$w<F<2w$, where $F=1/2-2V$ or $4V-1/2$, respectively.
Reducing the vote share of the candidate at $r$ to at most $w$ then requires a strict pair in $(1/2,r)$ or $(r,s)$,
respectively, also of capacity $F$; its other affecting gaps
and copies have capacity below $1/10$.
Together with the left pair and the new candidate in the exterior region this requires
5 new candidates, exceeding $m$.
At $w=V$, entry in the longer gap adjacent to $c$ is necessary.
That new candidate and all old candidates at positions in $\mathcal R$
win, accounting for the maximum of $m$ winners. An additional winner
would give the next player less than its guaranteed payoff $1/m$.
Thus each of the stated players receives exactly $1/m$; the argument
does not require a unique position for the new candidate on the left.
\end{proof}

\section{Case 3: 4 new candidates}\label{sec:case-three}
We treat the third case of Table~\ref{tab:improvement-families}
throughout this section. We first exclude the configurations
with coincident positions for $V>7/40$, then those with distinct
positions. We finally treat $V=7/40$, where player~1 also wins.

All histories considered in this section extend $(1/2,1/10)$.
Throughout this section, let $7/40\le V<1/5$ and define
\[
 a=1/2-2V,\quad b=1/10+2V,\quad r=3/2-4V,\quad s=1/2+2V.
\]
The new positions in Case 3 are $\{a,b,r,s\}$ or $\{a,b,r,r\}$.
For the distinct-position case, let
\[
 \mathcal C=\{1/10,a,b,1/2,r,s\}
\]
denote the configuration with 1 candidate at each listed position.
The candidates at $a,b,r,s$ have vote share $V$.
Players~2 and~1 have vote shares $\alpha=3/10-V$ and
$\beta=7/10-3V$: both lose for $V>7/40$, while player~1 also wins
at $V=7/40$. Put $E=1/2-2V$ and $q=V/2$. The relevant bounds are
\[
  1/10<\alpha,\quad \beta\le V,\quad q<1/10<E\le2V-1/5<V,
  \qquad 3V-1/2,\ V-1/10,\ 3V/2-1/5<q.
\]
The gaps $(a,b),(1/2,r),(r,s)$ have capacities $2V-1/5,E,3V-1/2$;
the supremum of entry capacities to the right of $s$ is $E$.
The gaps $(1/10,a),(b,1/2)$ have capacity $1/5-V<1/12$.

\subsection{Case 3: a copied rightmost position}\label{sec:case-three-copies}
\begin{lemma}\label{lem:case-three-copies}
Let $7/40<V<1/5$ and $a_3\in[0,1]\cup\{\OUT\}$.
No SPNE of $\Gamma(1/2,1/10,a_3)$ has an outcome in which
exactly 4 players enter after $(1/2,1/10)$, at positions
forming the multiset $\{a,b,r,r\}$.
\end{lemma}
\begin{proof}
If player~12 enters at $a$ or $b$, moving slightly toward the
other of these positions strictly increases its payoff by
Lemma~\ref{lem:deletion}: the candidate on the other side is
player~2 or~1 and loses.
If player~12 chooses $\OUT$ or enters at $r$, then
$1-r=4V-1/2>V$. Lemma~\ref{lem:copied-extreme-response}
gives a profitable deviation that makes player~12 the sole winner.
Thus no SPNE induces this configuration for $V>7/40$.
\end{proof}
The case $V=7/40$ is treated in Section~\ref{app:boundaries}.

\subsection{Case 3: distinct positions and the order of entry}\label{sf:four-timing}
For $V>7/40$, players~1 and~2 lose in $\mathcal C$.
We exclude this outcome in every SPNE. The auxiliary lemmas also
cover $V=7/40$, since they will be used in Section~\ref{app:boundaries}.

\begin{proposition}\label{sf:main}
Fix $a_3\in[0,1]\cup\{\OUT\}$.
No SPNE of $\Gamma(1/2,1/10,a_3)$ induces the distinct-position
configuration $\mathcal C$ of Case 3 with $7/40<V<1/5$.
\end{proposition}

First, we show that certain changes of position remain profitable
when at most 5 players remain. With 6 players remaining, we identify
which existing candidates win and prove that further entry occurs.
These results restrict the order of entry and the number of players
left after the later entry in $(1/10,1/2)$.
A final deviation to $s$ then increases that player's payoff from
$1/4$ to $1/3$ in every SPNE after the deviation.

For $V>7/40$, consider replacing the action $a$ by $3/20$, $b$ by
$9/20$, or $r$ by $4/5$, holding all other entry actions fixed.
Each change preserves the deviator's vote share $V$ and makes the
candidate at $b$, $a$, or $s$, respectively, lose, leaving exactly
3 winners. Call the resulting configurations the 3 modified configurations.

\begin{lemma}\label{lem:four-entry-bounds}\label{sf:four-timing-1}
Let $h$ extend $(1/2,1/10)$ and have configuration $\mathcal C$ or,
when $V>7/40$, one of the 3 modified configurations.
If at most 5 players remain, every SPNE of $\Gamma(h)$ induces
an outcome in which all remaining players choose $\OUT$.
If $h$ has configuration $\mathcal C$ and exactly 6 players remain, the
maximal vote share in any SPNE outcome with further entry is
at most $1/10$.
\end{lemma}
\begin{proof}
Each capacity below is evaluated at the history under consideration.
References to a gap after a change mean the gap between the
corresponding candidates at their resulting positions.

\paragraph{Low shares require 6 new candidates.}
Fix a history with configuration $\mathcal C$ or one of the 3 modified
configurations. Old and new candidates in this calculation are
defined relative to this history. Consider a terminal extension in
which at least 1 remaining player enters and every such player wins;
let $w$ be its maximal vote share.
The gaps corresponding to $(1/10,a),(b,1/2)$ have entry capacities
below $1/12$, so no winning new candidate can enter there.
Their midpoints therefore delimit 3 fixed intervals that remain
unions of vote cells, containing 1, 2 and 3 old candidates.
Their lengths are
respectively $\alpha$ or $1/8$, $2V$ or $V+7/40$, and
$7/10-V$ or $21/40$: these exceed $1/10,3/10,1/2$.
At $w\le1/10$, Lemma~\ref{lem:floor} requires at least
1, 2 and 3 new candidates, respectively. Thus at least 6 players
must enter, including players who enter at the positions of existing candidates.

\paragraph{Possible terminal vote shares above $1/10$.}
Fix a history with configuration $\mathcal C$ and at most 6 players remaining,
or one of the modified configurations and at most 5 players remaining.
Again consider a terminal extension with at least 1 new candidate,
in which every new candidate wins, now with maximal vote share $w>1/10$.
Every entry capacity is below $V$. All interior gaps other than the
2 corresponding to $(a,b),(1/2,r)$, and all copies, have capacity
below $1/10$. The capacities of these 2 gaps are at least $E$:
replacing $a$ or $b$ changes the left one to $V-1/40$; replacing $r$
changes the right one to $3/20$.

At least 1 new candidate must enter strictly to the right of $s$.
With configuration $\mathcal C$, or after changing the action at $a$ or $b$,
the candidate at $s$ would otherwise retain vote share $V$,
which exceeds every entry capacity. After the candidate at $r$ moves to $4/5$, the
candidate at $s$ instead retains vote share $7/20-V>3/20$.
This would force $w\ge7/20-V$, leaving the deviator's vote share
equal to $V$, which exceeds every entry capacity.
Thus $w<E$. Each of the 2 gaps just identified is bordered by
a candidate with vote share $V$. Reducing that candidate's vote
share to at most $w$ requires entry in that gap. Both gaps must therefore contain new candidates. Lemma~\ref{lem:short-gaps} gives 2 strict pairs, and
only 1 new candidate fits in the exterior region because $E<2w$.
Each old neighbor of a new candidate then has vote share at most
$V-w<w$. Player~2 may tie, but the deviation in Lemma~\ref{gl:spare-deletion} does not change that player's vote share.
All other interior gaps and the left exterior region have entry capacities below $w$.
Lemma~\ref{gl:spare-deletion} excludes these configurations as SPNE
outcomes: exactly 5 players enter after the history under consideration,
so at most $6-5=1$ player moves after the last of these entries. Thus any SPNE with further entry from $\mathcal C$
with 6 players remaining has maximal vote share at most $1/10$.
With at most 5 players remaining, the low-share count also excludes
further entry, both from $\mathcal C$ and from each modified configuration.
All remaining players therefore choose $\OUT$ in every such SPNE.
\end{proof}

For $V>7/40$, each of the 3 changes therefore gives the deviator
payoff $1/3>1/4$ in every SPNE after the deviation when at most
5 players remain.

To determine whether the last of the 4 new candidates could be
player~6, we must also consider configuration $\mathcal C$ with 6 players
remaining. We first identify the winners. A separate payoff guarantee
will then establish further entry.

\begin{lemma}\label{sf:four-retention}
Let $h$ extend $(1/2,1/10)$, have configuration $\mathcal C$, and leave
exactly 6 players to move. In the outcome of every SPNE of
$\Gamma(h)$, the candidates at $a,b,r,s$ win; player~1 also wins
if $V=7/40$. If such an outcome includes entry after $h$, all
6 remaining players enter and the maximal vote share is $q=V/2$.
\end{lemma}
\begin{proof}
Fix an SPNE of $\Gamma(h)$.
If all remaining players choose $\OUT$, the winning claims follow
from the vote shares in $\mathcal C$. Otherwise, Lemma~\ref{lem:allwin}
makes every player entering after $h$ a winner, and
Lemma~\ref{lem:four-entry-bounds} gives maximal vote share $w\le1/10$.
The regional bounds in its proof require all 6 players to enter:
1, 2 and 3 in the 3 fixed regions, respectively.
The region containing the candidates at $a,b$ has length $2V$
and contains 4 candidates in total, so $w\ge q$.
For $q<w\le1/10$, entry at any old candidate's position has capacity
at most $q<w$. The new candidate in the leftmost region must
therefore enter strictly to the left of $1/10$, implying $w<1/10$.
The 2 new candidates in the region containing $a,b$ enter in $(a,b)$: its half-length $2V-1/5$ lies in
$(w,2w)$, so they form a strict pair. Reducing the vote share of the candidate at $r$ to at most $w$
requires entry in $(1/2,r)$, since the capacity of $(r,s)$ and its copy capacity are
below $w$. The half-length $E$ of $(1/2,r)$ also lies in $(w,2w)$,
forcing a second strict pair. The remaining new candidate on the right must
enter strictly to the right of $s$, since entry in $(r,s)$ or at $s$
has capacity below $w$.
Thus there are exactly 2 strict pairs and 2 new candidates in the exterior regions.
The old vote shares are $\alpha-w$, $\beta-w$, and, at the other 4 positions,
$V-w$, all below $w$. Both new candidates in the exterior regions are in $(0,1)$.
All 6 remaining players enter, so the last of them is player~12.
If that player enters in an interior gap, moving slightly toward
the other new candidate in that gap increases its payoff by
Lemma~\ref{lem:deletion}. If it enters in an exterior region,
moving slightly toward the nearest candidate makes it the sole winner.
No player moves after this deviation, so $q<w\le1/10$ is impossible
in an SPNE.

At $w=q$, the region containing $a,b$ has length $2V=4q$ and
contains 4 candidates. All 4 therefore win, including the old
candidates at $a,b$. For $V>7/40$, player~1
has vote share $\beta>q$ and copy capacity below $q$, so 2 new
candidates at distinct positions in $(1/2,r)$ are necessary.
The pair reduces the vote share at $r$ from $2q$ to $q$. The remaining new
candidate on the right copies $s$ or enters beyond it: after the
pair in $(1/2,r)$, the capacities of $(r,s)$ and of copies at
$1/2$ or $r$ are below $q$. Either action leaves the old candidate at $s$ with vote share $q$. At $V=7/40$,
the rightmost fixed region has length $6q$ and contains 6 candidates.
All 6 win, including the candidates at $1/2,r,s$.
Thus, in every SPNE of a subgame with configuration $\mathcal C$ and
6 players remaining, the candidates at $a,b,r,s$ win. Player~1
also wins when $V=7/40$. Whenever further entry occurs, all 6
remaining players enter and the maximal vote share is $q$.
\end{proof}

\begin{lemma}\label{sf:four-timing-2}\label{sf:four-timing-2-1}\label{sf:four-timing-2-2}\label{sf:four-timing-2-3}\label{sf:four-timing-2-4}
Let $h$ extend $(1/2,1/10)$, have configuration $\mathcal C$, and leave
exactly 6 players to move. Choosing $r-V$ guarantees the next
player a payoff of at least $1/12$. Consequently, that player
enters in every SPNE of $\Gamma(h)$.
\end{lemma}
\begin{proof}
Consider the following 6 successive entry actions:
\[
 x_1=r-V,\quad l=V-1/10,\quad x_2=1/10+V,\quad s,
 \quad 1/2-V,\quad 1/2+V.
\]
The fourth action places a second candidate at $s$.
Let $h_0=h$ and let $h_i$ be the history obtained by appending
the first $i$ actions in this sequence to $h$, for $1\le i\le6$.
Let $P_i$ be the set of players who took these $i$ actions,
with $P_0=\varnothing$. We apply Lemma~\ref{lem:positive-chain}.
For each $1\le i<6$, consider a terminal extension of $h_i$ in which
at least 1 remaining player enters and every such player wins;
let $w$ denote its maximal vote share.
After entry at $x_1$, player~1 retains vote share
$\beta'=7/10-7V/2$: its adjacent gap and copy capacities are below
$1/12$. For $w\le q$, before entry at $l$, entry in the gap immediately to the right of
player~2 or at its position has capacity below $1/12$. Reducing
player~2's vote share to at most $w$ requires 1 new candidate
strictly to the left of $1/10$ (2 vote shares of $w\ge1/12$ exceed the mass $1/10$ to the left of $1/10$), leaving vote share $\alpha-w$. After entry at $l$, player~2 retains vote share
$\alpha-q\le\alpha-w$; its adjacent gap and copy capacities are
below $1/12$. At most 10 other candidates have vote shares at most $w$,
so
\[
 1\le\alpha-w+\beta'+10w=1+9(w-q)\le1.
\]
Thus $w=q$, and every candidate other than players~2 and~1 wins,
including every new candidate in the specified sequence.

For $w>q$, before entry at $x_2$ the candidate at $x_1$ keeps vote
share $E$, because its affecting capacities are at most $q$.
Then $w\ge E$ keeps the vote share of the candidate at $s$ fixed at $V$,
above every entry capacity. After entry at $x_2$ and before the last action in the sequence,
the candidate at $x_1$ keeps vote share $E$, and every entry capacity
is below $E$: all interior and copy capacities are at most $q$,
and the exterior supremum $E$ is unattained.
Hence no terminal extension just described can have maximal vote
share above $q$.
In the final configuration, the candidates at $a,b,r,s$ and the 6 prescribed new candidates
have vote share $q$. Player~2 has $\alpha-q<q$; player~1 has $\beta'<q$
for $V>7/40$ and $\beta'=q$ at $V=7/40$. Thus each prescribed
new candidate receives $1/10$ or $1/11$, respectively, and hence at least $1/12$.
Lemma~\ref{lem:positive-chain} therefore shows that choosing $x_1$
guarantees the next player a payoff of at least $1/12$.
Consequently, that player enters in every SPNE of $\Gamma(h_0)$.
\end{proof}

\begin{corollary}\label{sf:four-exclusion}\label{sf:four-timing-3}
Let $7/40<V<1/5$ and fix $a_3\in[0,1]\cup\{\OUT\}$.
Suppose an SPNE of $\Gamma(1/2,1/10,a_3)$ induces configuration $\mathcal C$.
Among players~3--12 who enter, the first chooses $r$ and the last
chooses $s$. The later of the 2 players choosing $a$ and $b$ has
index $j$ with $6\le j\le11$, so 1 to 6 players move after it.
\end{corollary}
\begin{proof}
Suppose an SPNE has the proposed strict outcome with 4 new candidates.
Lemma~\ref{lem:left-interval-response} excludes $a,b$ as the first
new entry: the next player could guarantee payoff 1, which would
make the first new candidate lose. The last new candidate cannot
be player~6: its entry would leave configuration $\mathcal C$ with 6 players
remaining, and Lemma~\ref{sf:four-timing-2} forces further entry.
The last new candidate is therefore player~7 or later. At most
5 players remain after its action, so Lemma~\ref{lem:four-entry-bounds}
applied to the 3 deviations above excludes
$a,b,r$ as its position. Thus the first new candidate chooses $r$
and the last chooses $s$.

The later of the 2 players entering in $(1/10,1/2)$ cannot be
player~5. If it were, player~6 could choose $s$, reaching configuration
$\mathcal C$ with 6 players remaining. Lemma~\ref{sf:four-retention} guarantees that
the candidate at $s$ wins in every SPNE of that subgame, so this
action gives player~6 a positive payoff. Player~6 must therefore
enter in every SPNE. In the proposed outcome its only remaining
position is $s$, but Lemma~\ref{sf:four-timing-2} then forces
at least 1 subsequent player to enter, a contradiction.
Hence the later player entering on the left is player~6 or later.
Since entry at $s$ follows it, its index lies between 6 and 11,
leaving 1 to 6 players after its deviation.
\end{proof}

\begin{proof}[Proof of Proposition~\ref{sf:main}]
Suppose an SPNE of $\Gamma(1/2,1/10,a_3)$ induces
a strict configuration with 4 new candidates, using the positions
from the beginning of Section~\ref{sec:case-three}. Let $j$ be the later of the
2 players entering in $(1/10,1/2)$, and let $c\in\{a,b\}$ be the
position chosen by the earlier one.
By Corollary~\ref{sf:four-exclusion}, $r$ has already been chosen,
$s$ has not, and $6\le j\le11$.
If player~$j$ chooses $s$ instead, the resulting configuration is
$\{1/10,c,1/2,r,s\}$, with 1 to 6 players remaining.
The next player can choose $3/20$ if $c=b$, or $9/20$ if $c=a$.
The resulting configuration is one of the modified configurations
in Lemma~\ref{lem:four-entry-bounds}, with at most 5 players remaining.
That lemma forces all subsequent players to choose $\OUT$ and
gives the next player payoff $1/3$.
The hypothesis of Lemma~\ref{lem:left-pair-completion} is therefore
satisfied with $\mathcal R=\{r,s\}$ and $m=3$.
It gives the deviator, now the candidate at $s$, payoff $1/3$ in every
SPNE after the deviation. This exceeds its original payoff $1/4$.

\end{proof}

\subsection{\texorpdfstring{Case 3 at $V=7/40$: player 1 also wins}{Case 3 at V=7/40: player 1 also wins}}\label{app:boundaries}
New candidates in this subsection are defined relative to $(1/2,1/10)$,
unless an auxiliary history is explicitly specified. We consider the
2 remaining multisets of positions chosen by players~3--12 who enter:
\[
 \{3/20,9/20,4/5,17/20\}\qquad\text{or}\qquad\{3/20,9/20,4/5,4/5\}.
\]
The term \emph{boundary configuration} below refers to one of these 2
configurations, corresponding to $V=7/40$ in Case 3 of
Table~\ref{tab:improvement-families}.
In each configuration, player~1 and the 4 new candidates win with
vote share $7/40$ and payoff $1/5$, while player~2 loses. Thus the
exclusion in Proposition~\ref{sf:main} does not apply.
Player~3's action must belong to $\{3/20,9/20,4/5,17/20\}$ for either
multiset to occur. At the first 2 actions, player~4 can guarantee payoff 1,
which excludes both outcomes. At $4/5$, we choose a full SPNE strategy
profile avoiding both multisets. At $17/20$, we first show that an SPNE
with the first multiset would require player~4 to choose $\OUT$,
then establish a positive payoff guarantee contradicting that choice.

\begin{proposition}\label{bd:excluded}
Fix $a_3\in[0,1]$ in $\Gamma(1/2,1/10)$. The subgame
$\Gamma(1/2,1/10,a_3)$ has an SPNE in which the multiset of
positions chosen by candidates among players~3--12 is neither
$\{3/20,9/20,4/5,17/20\}$ nor $\{3/20,9/20,4/5,4/5\}$.
For $a_3\ne4/5$, neither multiset occurs in any SPNE of this subgame.
For $a_3=4/5$, such an SPNE can be chosen in which at least 1 of
players~4--12 enters.
\end{proposition}

\subsubsection{\texorpdfstring{$a_3=3/20$ or $9/20$: player 4 can guarantee payoff 1}{a3=3/20 or 9/20: player 4 can guarantee payoff 1}}\label{bd:case-left}
\begin{corollary}\label{cor:boundary-left}
Let $a_3\in\{3/20,9/20\}$. In $\Gamma(1/2,1/10,a_3)$,
action $29/40$ guarantees player~4 payoff 1.
Consequently, player~4 is the sole winner in every SPNE of this subgame,
and neither boundary configuration occurs.
\end{corollary}
\begin{proof}
For either value of $a_3$, Lemma~\ref{lem:left-interval-response}
gives $z=29/40$ and the stated guarantee. Optimality gives player~4
payoff 1 in every SPNE. Player~3 therefore loses, whereas it wins
in both boundary configurations.
\end{proof}

\subsubsection{\texorpdfstring{$\Gamma(1/2,1/10,4/5)$: player 4 can enter at $9/10$}{Gamma(1/2,1/10,4/5): player 4 can enter at 9/10}}\label{bd:case-four-fifths}
We first establish a payoff guarantee for player~4, then use it to
choose a full SPNE of $\Gamma(1/2,1/10,4/5)$ that avoids both
specified multisets and has entry by at least 1 of players~4--12.

\begin{lemma}\label{bd:right-response-R}
In $\Gamma(1/2,1/10,4/5)$, action $9/10$ guarantees player~4 a payoff
of at least $1/5$: every SPNE of
$\Gamma(1/2,1/10,4/5,9/10)$ gives that player at least $1/5$.
\end{lemma}
\begin{proof}
Set
\[
 \mathcal G_0=\{1/10,1/2,4/5,9/10\},\quad \mathcal G_1=\mathcal G_0\cup\{2/5\},\quad
 \mathcal G_2=\mathcal G_1\cup\{1/5\},\quad \mathcal G_3=\mathcal G_2\cup\{13/20\}.
\]
Let $h_0=(1/2,1/10,4/5,9/10)$ and let
$h_1=(h_0,2/5)$, $h_2=(h_1,1/5)$, and $h_3=(h_2,13/20)$.
History $h_i$ has configuration $\mathcal G_i$; the respective
numbers of remaining players are 8, 7, 6, and 5.
For each $i=0,1,2$, consider a terminal extension of $h_i$ in which
at least 1 player enters after $h_i$ and every such player wins.
Old and new candidates in this calculation are defined relative
to $h_i$. Let $k\ge1/12$ be the maximal vote share.
If $k<1/10$, the candidate at $4/5$ has vote share $1/5$ at $h_i$.
Its adjacent interior gaps have half-lengths $3/20\in(k,2k)$
and $1/20<k$. Lemma~\ref{lem:two-sided-repair} would require
2 adjacent gaps containing new candidates, whereas only 1 can
contain a winning new candidate.
For $k\ge1/10$, the candidate at $9/10$ retains vote share $3/20$.
Entry in $(4/5,9/10)$ has capacity $1/20$, a copy at $9/10$
has capacity $3/40$, and every entry strictly to the right of
$9/10$ has capacity below $1/10$. Lemma~\ref{lem:protection}
therefore applies, and $k\ge3/20$.

At $h_1,h_2$, every entry capacity is at most $3/20$, so $k=3/20$.
At $h_0$, a maximal vote share $k>3/20$ would leave the candidate
at $4/5$ with vote share $1/5$, giving $k\ge1/5$.
All entry capacities are at most $1/5$, so $k=1/5$.
The only possible winning new candidate would then be a singleton
at a position $y\in(1/10,1/2)$. The vote-share bounds at the old
positions $1/10$ and $1/2$ would require, respectively, $y\le3/10$
and $y\ge2/5$, a contradiction.
Thus every terminal extension under consideration has $k=3/20$.
At $h_0$, the candidate at $9/10$ therefore wins in every such extension.

At $h_1$, reducing the old candidates' vote shares to at most $k$
requires 1 new candidate at a position $y\in(1/10,2/5)$.
Entry at $1/10$ or $2/5$ has capacity $1/8$ or $1/10$, respectively,
both below $k$.
The vote-share bounds at $1/10$ and $2/5$ give $y\le1/5$ and
$y\ge1/5$, respectively. Hence $y=1/5$, and the old candidate
at $2/5$ has vote share $3/20$ and wins.
At $h_2$, the candidate at $1/5$ retains vote share $3/20$, since
every entry affecting it has capacity at most $1/10<k$.

At $h_3$, every entry capacity is at most $1/10$, whereas every
entry affecting the candidate at $13/20$ has capacity $3/40<1/12$.
By Lemma~\ref{lem:protection}, that candidate retains vote share
$3/20$ in every SPNE outcome. Any player entering after $h_3$
would have final vote share at most $1/10$ and would lose.
Lemma~\ref{lem:allwin} therefore implies that all remaining players
choose $\OUT$ in every SPNE of $\Gamma(h_3)$. The candidates at
$1/10,1/5,2/5,13/20,9/10$ win with vote share $3/20$ and payoff $1/5$;
the candidates at $1/2,4/5$ have vote share $1/8$.
Apply Lemma~\ref{lem:positive-chain} to $h_0,h_1,h_2,h_3$, with
$P_0=\{4\}$, $P_1=\{5\}$, $P_2=\{6\}$, $P_3=\{7\}$,
and $v=1/5$. Player~4, who entered at $9/10$, therefore
receives at least $1/5$ in every SPNE of
$\Gamma(1/2,1/10,4/5,9/10)$.
\end{proof}

\begin{corollary}\label{cor:boundary-R-selection}
The subgame $\Gamma(1/2,1/10,4/5)$ has an SPNE in which
player~4 enters and neither boundary configuration occurs.
\end{corollary}
\begin{proof}
Pure SPNE exist in every subgame by the backward-induction argument
in Section~\ref{sec:positive-continuation}. Fix an SPNE of
$\Gamma(1/2,1/10,4/5,9/10)$ and let $\rho$ be player~4's payoff
in it. Lemma~\ref{bd:right-response-R} gives $\rho\ge1/5$. Choose a full SPNE strategy profile in
the subgame following every other action of player~4, including $\OUT$.
Let player~4 choose a best response to these profiles, choosing $9/10$
whenever it is among the best responses. A best response exists
because the set of possible payoffs is finite.
If player~4 chooses $9/10$, its position belongs to neither of the
2 multisets. Otherwise its payoff exceeds $\rho\ge1/5$, whereas
in either specified configuration that player could receive at most
$1/5$. Thus the resulting SPNE avoids both configurations.
Player~4's payoff is positive, so that player enters.
\end{proof}

\subsubsection{\texorpdfstring{$\Gamma(1/2,1/10,17/20)$: player 4's incentive to enter}{Gamma(1/2,1/10,17/20): player 4's incentive to enter}}\label{bd:case-seventeen-twentieths}
There are 2 parts to this case. We first show that an SPNE whose
new candidates choose $\{3/20,9/20,4/5,17/20\}$ would require
player~4 to choose $\OUT$. We then prove that choosing $27/40$
gives that player a positive payoff in every SPNE of the resulting
subgame, contradicting the optimality of $\OUT$.

Here $V=7/40$, so
$\mathcal C=\{1/10,3/20,9/20,1/2,4/5,17/20\}$.
We first classify SPNE outcomes when 1 of the positions $9/20,4/5$
has not yet been chosen.

\begin{lemma}\label{lem:boundary-missing-classification}
Let $h$ extend $(1/2,1/10,17/20)$, have configuration
$\mathcal C\setminus\{9/20\}$ or $\mathcal C\setminus\{4/5\}$, with exactly 1
candidate at each listed position and no others, and leave at most
7 players to move. In every SPNE of $\Gamma(h)$ whose outcome
includes entry after $h$, one of the following holds:
\begin{enumerate}[label=(\roman*),nosep]
\item The maximal vote share is $7/80$, all 12 players enter,
and every candidate except player~2 wins.
\item Exactly 1 player enters after $h$, at the missing position in
$\mathcal C$, and the maximal vote share is $7/40$.
\end{enumerate}
\end{lemma}
\begin{proof}
Consider a terminal extension of $h$ in which at least 1 player
enters after $h$ and every such player wins. Every SPNE outcome
with further entry has these properties by Lemma~\ref{lem:allwin}.
Old and new candidates in this proof are defined relative to $h$.
Let $k\ge1/12$ be the maximal vote share in this extension.
The 2 longer interior gaps are $(3/20,1/2),(1/2,4/5)$ in the first case and $(3/20,9/20),(1/2,17/20)$ in the
second, with half-lengths $7/40,3/20$ in opposite orders.
Every other gap has half-length $1/40$.

For $k<1/10$, reducing the vote share of the candidate at $1/10$
from $1/8$ to at most $k$ requires entry strictly to its left:
all other entries affecting it have capacity below $1/12$.
Exactly 1 new candidate enters there, since 2 winning candidates
would have total vote share at least $1/6$, exceeding the length
$1/10$ of the left exterior region. The vote cells of this new
candidate and the candidate at $1/10$ together form the fixed
interval $[0,1/8]$. The other candidates, at most 10 in number,
therefore have total vote share $7/8$, giving $k\ge7/80$.
Equality requires 12 candidates in total and makes every candidate
except the one at $1/10$ a winner. For $7/80<k<1/10$,
Lemma~\ref{lem:two-sided-repair} applied to the candidate at $3/20$
(when $9/20$ is missing) or $1/2$ (when $4/5$ is missing) gives a
contradiction: its vote share is $1/5$, with adjacent half-gaps $1/40,7/40$.

For $k\ge1/10$, every entry affecting the candidate at $1/10$
has capacity below $k$. That candidate retains vote share $1/8$
by Lemma~\ref{lem:protection}, so $k\ge1/8$.
If $k<3/20$, each long gap is bordered by a candidate with vote
share $7/40$ or $1/5$. Entry in that candidate's other adjacent
gap and a copy at its position both have capacity below $k$.
Each gap therefore contains exactly 2 new candidates at distinct
positions, by Lemma~\ref{lem:short-gaps}. Every old cell is then at
most $13/40-k<2k$, excluding copies. Without entry strictly to the right of $17/20$, the candidate there
would still have vote share at least $7/40>k$. Entry in its adjacent
interior gap has capacity below $k$ after the 2 pairs are specified,
so exactly 1 new candidate must be strictly to its right: 2 such
candidates with vote share $k$ would exceed the exterior-region
length $3/20<2k$.
Thus the terminal configuration has exactly 5 new candidates.
Every entry capacity in that configuration is below $k$.
Every old candidate adjacent to one of these 5 new positions loses: its final vote share is at most
$\max\{1/5-k,13/40-2k\}<k$. The candidate at $1/10$ may tie at $k=1/8$
but is not such a neighbor. All other interior gaps and the left exterior region have entry
capacities below $k$. Lemma~\ref{gl:spare-deletion} therefore excludes
this configuration as an SPNE outcome of $\Gamma(h)$: after the last
of the 5 new candidates enters, at most $7-5=2$ players remain.

At $k=3/20$, the required entries in the 2 longer gaps leave total
vote share $7/40$ to candidates at or to the right of $17/20$.
Lemma~\ref{lem:exterior-mass} would require exactly 1 new candidate
strictly to the right of $17/20$, but also requires their number
to be strictly less than $(3/20)/k=1$, a contradiction.
For $k>3/20$, the candidate at $17/20$ (when $9/20$ is missing)
or $3/20$ (when $4/5$ is missing) keeps vote share $7/40$.
Consequently $k=7/40$, since every entry capacity is at most $7/40$.
Only the gap of half-length $7/40$ can contain a winning new
candidate, and it contains exactly 1, at a position $x$.
If $9/20$ is missing, the vote-share bounds at $3/20$ and $1/2$
give $x\le9/20$ and $x\ge9/20$. If $4/5$ is missing, the bounds
at $1/2$ and $17/20$ give $x\le4/5$ and $x\ge4/5$.
Thus the new candidate chooses the missing position in $\mathcal C$.

It follows that every SPNE of $\Gamma(h)$ with further entry has
one of 2 types of outcome. Either the maximal vote share is $7/80$,
all 12 players have entered, and every old candidate except the one
at $1/10$ wins; or exactly 1 player enters after $h$, at the missing
position in $\mathcal C$, and the maximal vote share is $7/40$.
\end{proof}

\begin{corollary}\label{cor:boundary-missing-winners}
Under the hypotheses of Lemma~\ref{lem:boundary-missing-classification},
suppose exactly 7 players remain at $h$. The next player enters in
every SPNE of $\Gamma(h)$. The candidate at $3/20$ wins in every
such SPNE outcome; the candidate at $9/20$ also wins when the
configuration at $h$ is $\mathcal C\setminus\{4/5\}$.
\end{corollary}
\begin{proof}
The next player can enter at the missing position in $\mathcal C$, reaching
configuration $\mathcal C$ with 6 players remaining. Lemma~\ref{sf:four-retention},
at $V=7/40$, shows that this player wins in every SPNE of the resulting
subgame. The action therefore guarantees a positive payoff.
Consequently the next player enters in every SPNE of $\Gamma(h)$.
Lemma~\ref{lem:boundary-missing-classification} then makes the
stated old candidates winners in either type of outcome.
\end{proof}

\begin{lemma}\label{bd:missing}
If an SPNE of $\Gamma(1/2,1/10,17/20)$ has an outcome in which
the positions chosen by candidates among players~3--12 form the
multiset $\{3/20,9/20,4/5,17/20\}$, then player~4 chooses $\OUT$
on its equilibrium path.
\end{lemma}
\begin{proof}
Suppose such an SPNE has player~4 enter.
Its action must be in $\{3/20,9/20,4/5\}$.
Player~5 can choose $3/20$ if no candidate is there, and $9/20$
otherwise. The resulting history has configuration
$\mathcal C\setminus\{9/20\}$ or $\mathcal C\setminus\{4/5\}$, with exactly 7 players
remaining. Corollary~\ref{cor:boundary-missing-winners} guarantees
that player~5 wins in every SPNE after this action. Hence player~5 enters in every
SPNE of its subgame, although its optimal action need not be the
one just described.

On the proposed equilibrium path, player~5 must enter at one of
the 2 remaining positions of the specified multiset. Player~6 can
then choose the last missing position and reach configuration $\mathcal C$
with 6 players remaining. Lemma~\ref{sf:four-retention} guarantees
that player~6 wins after this action, so player~6 must also enter.
In the proposed outcome it must choose that last missing position.
The positive entry guarantee in Lemma~\ref{sf:four-timing-2}
then forces a further player to enter, contradicting the specified
multiset. Therefore such an SPNE would require player~4 to choose $\OUT$.
\end{proof}

Lemma~\ref{bd:missing} requires player~4 to choose $\OUT$ in the
proposed outcome. The following payoff guarantee will contradict
that choice.

\begin{proposition}\label{bd:right-response-S}\label{bd:s975}
In $\Gamma(1/2,1/10,17/20)$, action $27/40$ guarantees player~4 a
positive payoff: every SPNE of
$\Gamma(1/2,1/10,17/20,27/40)$ gives that player a positive payoff.
\end{proposition}
Put $h_0=(1/2,1/10,17/20,27/40)$. We prove the guarantee in
2 steps: identify the winners when further entry occurs, and then
establish that further entry occurs in every SPNE of $\Gamma(h_0)$.

\begin{lemma}\label{lem:boundary-S-retention}
Consider a terminal extension of $h_0$ in which at least 1 of
players~5--12 enters and every such player is a winner.
Then all 12 players enter, the maximal vote share is $7/80$,
and every candidate except player~2 is a winner.
\end{lemma}
\begin{proof}
The candidates at
$1/10,1/2,27/40,17/20$ have vote shares $3/10,23/80,7/40,19/80$,
respectively. Consider a terminal extension of $h_0$ in which at
least 1 of players~5--12 enters and every such player wins.
Let $k$ be the maximal vote share. Lemma~\ref{lem:floor} gives
$k\ge1/12$.

If $k>7/80$, entry in either interior gap adjacent to $27/40$,
or a copy at $27/40$, has capacity $7/80<k$ at $h_0$.
By Lemma~\ref{lem:protection}, player~4 retains vote share $7/40$,
so $k\ge7/40$. Every entry affecting the candidate at $17/20$
also has capacity below $k$: the adjacent-gap capacity is $7/80$,
the copy capacity is $19/160$, and entry strictly to its right
has capacity below $3/20$.
That candidate therefore retains vote share $19/80$.
Every entry capacity at $h_0$ is at most $1/5<19/80$, so no new
candidate could win. This contradicts the required further entry.

For $1/12\le k\le 7/80$, Lemma~\ref{lem:two-anchor-region} gives
the candidate at $1/10$ vote share $3/10-3k$. At most 11 other candidates each
receive at most $k$, so
\[
 1\le3/10-3k+11k=3/10+8k\le1.
\]
Equality forces $k=7/80$, entry by all 12 players, and every
candidate except the candidate at $1/10$ to win, including the candidate at $27/40$.
\end{proof}

\begin{lemma}\label{lem:boundary-S-player-five}
In $\Gamma(h_0)$, action $3/20$ guarantees player~5 a payoff
of at least $1/12$. Consequently, player~5 enters in every SPNE
of $\Gamma(h_0)$.
\end{lemma}
\begin{proof}
Consider the following 8 successive entry actions:
\[
 \begin{gathered}
 3/20,\quad 3/40,\quad 13/40,\quad 39/40,\\
 11/40,\quad 9/20,\quad 5/8,\quad 4/5.
 \end{gathered}
\]
Let $h_i$ be the history obtained from $h_0$ by the first $i$
actions in this sequence, for $1\le i\le8$.
For each $1\le i<8$, consider a terminal extension of $h_i$ in which
at least 1 player enters after $h_i$ and every such player wins.
Let $k\ge1/12$ be the maximal vote share. First suppose $k\le7/80$.
At $h_1$, the candidate at $1/10$ has vote share $1/8$.
Entry in its adjacent interior gap or a copy at its position has
capacity below $1/12$. At least 1 new candidate must therefore enter strictly to its left.
There cannot be 2: their total vote share would be at least $1/6$,
exceeding the length $1/10$ of that exterior region.
The unique new candidate there chooses $2k-1/10$, leaving the
candidate at $1/10$ with vote share $1/8-k$.
After the prescribed entry at $3/40$, the candidate at $1/10$
retains vote share $3/80\le1/8-k$, since every entry affecting it
has capacity below $1/12$. With at most 11 other candidates,
\[
 1\le1/8-k+11k\le1.
\]
Thus $k=7/80$ and every candidate except the one at $1/10$ wins,
including each player who has already entered in the specified sequence.
For $k>7/80$, before entry at $13/40$ the protection argument
from the proof of Lemma~\ref{lem:boundary-S-retention} still applies
to the candidates at $27/40,17/20$ and gives a contradiction.
After entry at $13/40$ and before entry at $39/40$, the candidate
at $27/40$ would still retain vote share $7/40$, while every entry
capacity is below $7/40$. After entry at $39/40$, every entry
capacity is at most $7/80$. Thus no terminal extension under
consideration has $k>7/80$.
At the end, all 12 players have entered: the candidate at $1/10$
has vote share $3/80$, and the other 11 have vote share $7/80$ and payoff
$1/11$. In particular, every prescribed new candidate receives at least $1/12$.
Apply Lemma~\ref{lem:positive-chain} to $h_0,\ldots,h_8$, with
$P_0=\varnothing$, $P_i=\{5,\ldots,4+i\}$ for $1\le i\le8$,
and $v=1/12$. It follows that action $3/20$ guarantees player~5
a payoff of at least $1/12$ in $\Gamma(h_0)$.
Hence player~5 enters in every SPNE of this subgame.
\end{proof}

\begin{proof}[Proof of Proposition~\ref{bd:right-response-S}]
Lemma~\ref{lem:boundary-S-player-five} makes player~5 enter in
every SPNE of $\Gamma(h_0)$. By Lemma~\ref{lem:allwin}, every
player who enters after $h_0$ wins. Lemma~\ref{lem:boundary-S-retention}
therefore applies to each such SPNE outcome and makes player~4,
whose position is $27/40$, a winner. This proves the payoff guarantee.
\end{proof}

\begin{corollary}\label{cor:boundary-S-exclusion}
No SPNE of $\Gamma(1/2,1/10,17/20)$ induces either boundary configuration.
\end{corollary}
\begin{proof}
An SPNE of this subgame with new-candidate positions
$\{3/20,9/20,4/5,17/20\}$ would require player~4 to choose $\OUT$
by Lemma~\ref{bd:missing}. Proposition~\ref{bd:right-response-S}
gives that player a positive payoff guarantee, a contradiction.
The other specified multiset, $\{3/20,9/20,4/5,4/5\}$, cannot occur
because player~3 has already chosen $17/20$.
\end{proof}

\begin{proof}[Proof of Proposition~\ref{bd:excluded}]
Corollaries~\ref{cor:boundary-left} and~\ref{cor:boundary-S-exclusion}
exclude both specified multisets in every SPNE of
$\Gamma(1/2,1/10,a_3)$ for $a_3\in\{3/20,9/20,17/20\}$.
Corollary~\ref{cor:boundary-R-selection} supplies the required full
SPNE of $\Gamma(1/2,1/10,4/5)$, in which player~4 enters.
For every other $a_3\in[0,1]$, neither multiset contains $a_3$,
so neither can be the multiset of positions chosen by player~3
and subsequent candidates.
\end{proof}

In the chosen SPNE of $\Gamma(1/2,1/10,4/5)$, player~4 enters,
so player~3 is not the only candidate to enter after $(1/2,1/10)$.
Section~\ref{app:solo-outer} treats the other third-player actions for which
an outcome with no entry by players~4--12 remains to be excluded.

\subsection{Conclusion for Case 3}
Combining the preceding arguments gives the result for the entire
case, including both possible multisets and the case $V=7/40$.

\begin{proposition}\label{prop:case-three}
Fix $a_3\in[0,1]$.
The subgame $\Gamma(1/2,1/10,a_3)$ has an SPNE whose outcome
does not belong to Case 3 of Table~\ref{tab:improvement-families}.
If $a_3\ne4/5$, no SPNE of this subgame induces an outcome in Case 3.
If $a_3=4/5$, an SPNE avoiding Case 3 can be chosen in which
at least 1 of players~4--12 enters.
\end{proposition}
\begin{proof}
For $V>7/40$, Lemma~\ref{lem:case-three-copies} excludes the
copied-position configurations and Proposition~\ref{sf:main}
excludes the distinct-position configurations, in every SPNE.
At $V=7/40$, Proposition~\ref{bd:excluded} supplies each of the
stated conclusions, including further entry at $a_3=4/5$.
\end{proof}

\section{Case 4: 5 new candidates}\label{sec:case-four}
\label{sf:five-penultimate}\label{sf:five-upper}\label{sf:uniform-five}
We now treat Case 4 of Table~\ref{tab:improvement-families}.
All histories below extend $(1/2,1/10)$, and the number of remaining
players includes those who subsequently choose $\OUT$.

Let $3/20<V<1/6$ and write
\[\begin{gathered}
 a=1/2-2V,\qquad b=1/10+2V,\qquad r=3/2-6V,\\
 s=1/2+2V,\qquad t=3/2-4V.
\end{gathered}\]
\begin{proposition}\label{prop:case-four}
For every $a_3\in[0,1]\cup\{\OUT\}$, the subgame
$\Gamma(1/2,1/10,a_3)$ has an SPNE whose outcome does not belong
to Case 4 of Table~\ref{tab:improvement-families}.
\end{proposition}

We exclude coincident positions by final-player deviations.
For distinct positions, we first restrict the order of entry and
then prescribe optimal actions at intermediate histories.
These prescriptions make a strict configuration impossible on
the equilibrium path of the chosen SPNE profile.

\subsection{Case 4: coincident positions}\label{sec:case-four-copies}
The possible multisets are $\{a,b,r,s,s\}$ and $\{a,b,r,r,r\}$.
If player~12 enters at $a$ or $b$, it can move slightly toward
the other of these positions. In $\{a,b,r,s,s\}$, it can also
improve by moving from $r$ slightly toward $s$.
In each instance Lemma~\ref{lem:deletion} applies because the
candidate on the other side is player~1 or~2 and loses.

Otherwise player~12 chooses $\OUT$ or enters at the copied rightmost
position $z$. Its right exterior region has length
$1-z=1/2-2V$ or $6V-1/2$, respectively, both strictly above $V$.
Lemma~\ref{lem:copied-extreme-response} gives a strictly profitable
deviation making player~12 the sole winner.
Thus neither multiset occurs in any SPNE.

\subsection{Case 4: distinct positions and the order of entry}
\paragraph{Small deviations and subsequent entry.}
Start with 1 candidate at each position in
$\{1/10,a,b,1/2,r,s,t\}$ and no other candidates.
For $\delta>0$, consider the 4 changes
$a\mapsto a+\delta$, $b\mapsto b-\delta$,
$r\mapsto r+\delta$, and $t\mapsto t-\delta$,
each holding all other positions fixed.
For sufficiently small $\delta$, with no further entry,
the first 3 changes give the moving candidate payoff $1/4$,
and the last gives payoff 1.

\begin{lemma}\label{lem:five-no-further-entry}
For each $V\in(3/20,1/6)$ there exists $\delta_0>0$ such that
the following holds for every $0<\delta<\delta_0$.
If a history $h$ has one of the 4 modified configurations above
and at most 5 players remain, every SPNE of $\Gamma(h)$ induces
an outcome in which all remaining players choose $\OUT$.
\end{lemma}
\begin{proof}
Fix a sufficiently small $\delta>0$ and a history as in the statement.
Suppose a terminal extension has at least
1 remaining player enter and every such player win.
Let $w$ be the maximal vote share in that extension.
Before the deviation, the gaps $(1/10,a),(b,1/2),(1/2,r),(s,t)$
have entry capacities below $1/12$. Their midpoints delimit fixed
unions of vote cells containing, respectively, the candidates at
$1/10$; at $a,b$; at $r,s$; and at $t$. The cell of player~1 is
excluded from these 4 regions. Their lengths are
$\alpha=3/10-V$, $2V$, $2V$, and $V$, which strictly exceed
$1/10,3/10,3/10,1/10$. All these strict inequalities persist after
a sufficiently small deviation, with the gaps and midpoints
recomputed at the resulting positions. No winning subsequent
candidate can enter in these gaps. If $w<1/10$, Lemma~\ref{lem:floor}
requires at least 1, 2, 2, and 1 further candidates in the 4 regions,
respectively, exceeding the 5 remaining players.

For $w\ge1/10$, player~2 retains its vote share after the deviation:
entry in its adjacent interior gap or at its position has capacity
below $1/12$, and entry strictly to its left has capacity below
$1/10$. Before the deviation, its vote share exceeds the capacity
$2V-1/5$ of $(a,b)$ by
$\alpha-(2V-1/5)=1/2-3V>0$; this strict inequality persists after
a sufficiently small deviation. Thus $w$ exceeds every entry
capacity affecting the 2 candidates initially at $a,b$.
At least one of these candidates retains vote share $V$ after
the deviation, and no subsequent winning entry can change that share.
Every entry capacity is below $V$, so no terminal extension can
satisfy the stated entry and winning conditions.
Thus all remaining players choose $\OUT$ in every SPNE after each
deviation when at most 5 players remain.
\end{proof}

\begin{corollary}\label{cor:five-entry-order}
Fix $a_3\in[0,1]\cup\{\OUT\}$ and suppose an SPNE of
$\Gamma(1/2,1/10,a_3)$ induces the strict configuration
$\{1/10,a,b,1/2,r,s,t\}$, with $3/20<V<1/6$.
Among players~3--12 who enter, the first chooses $r$ or $t$,
and the last chooses $s$.
\end{corollary}
\begin{proof}
Suppose an SPNE induces the strict configuration with 5 new candidates.
At least 7 players have moved when the last of these candidates enters,
so at most 5 players remain. Lemma~\ref{lem:five-no-further-entry} and the payoffs computed above
exclude $a,b,r,t$ as its action. It must therefore choose $s$.

Lemma~\ref{lem:left-interval-response} excludes $a,b$ as the first
new entry, since the next player would obtain payoff 1 and the
first new candidate would lose. Since $s$ is chosen last, the
first of the 5 new candidates must choose $r$ or $t$.
\end{proof}

\subsection{Case 4: choosing compatible SPNE profiles}
The next result allows us to prescribe action $s$ before the
missing left position is chosen.

\begin{lemma}\label{lem:five-optimal-middle}
Let $\eta$ be a history with exactly 1 candidate at each of
$1/10,c,1/2,r,t$ and no other candidates, where $c\in\{a,b\}$.
Suppose 2 to 7 players remain, and let $i$ move next.
Every SPNE of $\Gamma(\eta,s)$ gives player~$i$ and the
candidates at $r,t$ payoff $1/4$.
For every action $x\in[0,1]\cup\{\OUT\}$, every SPNE of
$\Gamma(\eta,x)$ gives player~$i$ a payoff at most $1/4$.
Thus $s$ is optimal against every choice of SPNE strategy profiles
following player~$i$'s actions.
\end{lemma}
\begin{proof}
After action $s$, the history $(\eta,s)$ has configuration
$\{1/10,c,1/2,r,s,t\}$ and 1 to 6 players remaining.
The next player can enter sufficiently close to the missing left
position, toward $c$. Lemma~\ref{lem:five-no-further-entry} forces
all of the at most 5 subsequent players to choose $\OUT$.
This action therefore guarantees that next player payoff $1/4$.
Lemma~\ref{lem:left-pair-completion} now applies with
$\mathcal R=\{r,s,t\}$ and $m=4$.
Every SPNE of $\Gamma(\eta,s)$ consequently gives player~$i$,
now the candidate at $s$, and the candidates at $r,t$ payoff $1/4$.

We next show that no other action followed by an SPNE can give
player~$i$ more than $1/4$. Such a payoff would imply at most 3 winners
and hence at most 3 players entering after $\eta$, including $i$.
With 5 old candidates this gives maximal vote share $w>1/10$.
Every entry capacity at $\eta$ is at most $V$, so $w\le V$.
For $w<V$, at least 2 new candidates are needed in the longer left
gap. On the right, entry in $(1/2,r)$ cannot win, so the vote cells
of candidates at or to the right of $r$ retain total length
$1-(1/2+r)/2=3V$. There are 2 old candidates there, at $r,t$;
at least 2 more are needed, exceeding the bound of 3 new candidates.
For $w=V$, a new candidate must enter in $(r,t)$. Together with
$r,t$, this gives 3 candidates whose vote shares sum to $3V$,
so all 3 win. The necessary new candidate in the longer left gap
gives a fourth winner, again a contradiction.
Consequently, action $s$ is optimal against every choice of SPNE
strategy profiles following player~$i$'s actions.
\end{proof}

\begin{proof}[Proof of Proposition~\ref{prop:case-four}]
The coincident-position outcomes are excluded in every SPNE by
Section~\ref{sec:case-four-copies}. It remains to avoid the strict
configurations simultaneously.
For every $V\in(3/20,1/6)$, use the positions
of the strict configuration with 5 new candidates.
At every history satisfying the hypotheses of
Lemma~\ref{lem:five-optimal-middle},
choose a full SPNE strategy profile in which the next player chooses $s$.
That lemma proves this action optimal for every assignment
of SPNE strategy profiles to the subgames following that player's actions.
The positions $r,t$ of the candidates already on the right determine $V$ uniquely:
$V=(3/2-r)/6=(3/2-t)/4$, with
$r\in(1/2,3/5)$ and $t\in(5/6,9/10)$.
Thus the prescribed action $s=1/2+2V$ is uniquely determined at each
such history. These actions can be imposed simultaneously by backward
induction, including at histories with earlier $\OUT$ actions: at each
history the prescribed action remains optimal for every choice of
SPNE profiles in the subsequent subgames.

Suppose a profile satisfying these prescriptions induces a strict
configuration with 5 new candidates. Let $j$ be the later player
entering on the left, and let $c\in\{a,b\}$ be the position chosen
by the earlier one. Corollary~\ref{cor:five-entry-order} says that
the first new candidate chooses $r$ or $t$ and the last chooses $s$.
Thus, before player~$j$ moves, at least one of $r,t$ has been chosen,
and $s$ has not.

If both $r,t$ have been chosen, then $6\le j\le11$.
The history immediately before player~$j$ moves has configuration
$\{1/10,c,1/2,r,t\}$ and $13-j\in\{2,\ldots,7\}$ remaining players,
including $j$. The prescription from Lemma~\ref{lem:five-optimal-middle}
therefore makes player~$j$ choose $s$, contradicting its proposed
entry at the missing left position.

If exactly one of $r,t$ has been chosen, then $5\le j\le10$:
1 candidate on the left and 1 on the right have already entered
after player~2, while the missing right position and $s$ must still
be chosen after $j$ on the proposed path.
Player~$j$ can instead choose the missing position among $r,t$.
The resulting history has configuration $\{1/10,c,1/2,r,t\}$ and
$12-j\in\{2,\ldots,7\}$ remaining players, including those who
subsequently choose $\OUT$.
The prescribed next action is $s$. Lemma~\ref{lem:five-optimal-middle}
gives the deviator, now a candidate at $r$ or $t$, payoff $1/4>1/5$.
This excludes every strict configuration with 5 new candidates
under the simultaneously chosen SPNE profiles.
\end{proof}

\section{Case 1: player 3 is the only new candidate}
\label{app:solo-outer}
For each $a_3\in[1/2,4/5]$, consider $\Gamma(1/2,1/10,a_3)$.
Our goal is to choose a full SPNE strategy profile in each of these
subgames whose outcome includes entry by at least 1 of players~4--12.
The outcome to be avoided has all 9 remaining players choose $\OUT$.
We first exclude this outcome by profitable later entry on $[1/2,7/10]$
and at $4/5$. On $(7/10,4/5)$ we establish the existence of an SPNE in which a
further player enters.

\subsection{\texorpdfstring{Direct entry responses: $1/2\le a_3\le7/10$ and $a_3=4/5$}{Direct entry responses: 1/2<=a3<=7/10 and a3=4/5}}\label{app:triples}
\begin{lemma}\label{so:direct-entry}
For every $a_3\in[1/2,7/10]\cup\{4/5\}$, the outcome of every
SPNE of $\Gamma(1/2,1/10,a_3)$ includes entry by at least 1 of
players~4--12.
\end{lemma}
\begin{proof}
Suppose $1/2\le a_3<7/10$ and all players~4--12 choose $\OUT$.
Player~12 can instead enter at $7/10$ and win alone. Its vote
share is $2/5$ when $a_3=1/2$, and $13/20-a_3/2>3/10$
otherwise. After this deviation, every old candidate has vote
share at most $3/10$.

At $a_3=7/10$, player~4 can choose $9/10$. The resulting
configuration is the reflection of the 4-position configuration
in Lemma~\ref{at:grid-completion}, with 8 players remaining.
Player~4 at $9/10$ corresponds under reflection to the old
candidate at $1/10$ in that lemma. This action therefore
guarantees player~4 a payoff of at least $1/10$, so player~4
enters in every SPNE of $\Gamma(1/2,1/10,7/10)$.

At $a_3=4/5$, Lemma~\ref{bd:right-response-R} gives player~4
an action guaranteeing a payoff of at least $1/5$. Player~4
therefore enters in every SPNE of $\Gamma(1/2,1/10,4/5)$.
\end{proof}

At $a_3=4/5$, Section~\ref{app:boundaries} also provides a
choice of a full SPNE avoiding both configurations in Case 3
in which player~1 wins.

\subsection{\texorpdfstring{Possible new candidates for $7/10<a_3<4/5$}{Possible new candidates for 7/10<a3<4/5}}\label{so:inventory-section}
For the rest of this section, assume $7/10<a_3<4/5$ and put
\[
 q=(3/2-a_3)/8,\qquad 7/80<q<1/10.
\]
The goal on this interval is Corollary~\ref{so:further-entry}:
there exists a full SPNE of $\Gamma(1/2,1/10,a_3)$ with further entry.

\paragraph{The backward-induction argument.}
We compare the following actions at 4 successive histories:
\[
 r=a_3+2q,\qquad s=1/10+2q,\qquad x_0=a_3-2q,\qquad L=1/2-4q.
\]
Write $\mathcal C_5=\{1/10,s,1/2,a_3,r\}$ and
$\mathcal C_6=\mathcal C_5\cup\{x_0\}$.
Table~\ref{tab:solo-backward-route} lists the auxiliary configurations
reached by these successive actions, with no intervening $\OUT$.
The numbers in the middle column count all remaining players,
including those who might subsequently choose $\OUT$.
In each local argument below, old and new candidates are defined
relative to the history specified in that argument.
\begin{longtable}{@{}>{\raggedright\arraybackslash}p{0.52\textwidth}>{\centering\arraybackslash}p{0.22\textwidth}>{\centering\arraybackslash}p{\dimexpr0.26\textwidth-4\tabcolsep\relax}@{}}
\caption{Auxiliary histories for the construction on $7/10<a_3<4/5$.}\label{tab:solo-backward-route}\\
\toprule \textbf{Configuration} & \textbf{Decisions remaining} & \textbf{Action compared next}\\\midrule\endfirsthead
\toprule \textbf{Configuration} & \textbf{Decisions remaining} & \textbf{Action compared next}\\\midrule\endhead
$\{1/10,1/2,a_3\}$ & 9 & $r$\\
$\{1/10,1/2,a_3,r\}$ & 8 & $s$\\
$\mathcal C_5$ & 7 & $x_0$\\
$\mathcal C_6$ & 6 & $L$\\
$\mathcal C_6\cup\{L\}$ & 5 & ---\\\bottomrule
\end{longtable}

The proof works from the bottom row upwards.
First, every SPNE after the prescribed entry at $L$ gives the
candidates at $L,x_0,s$ payoff $1/10$ (Lemma~\ref{so:upper-base}). We show that, at the 2 preceding histories, players~7 and~6,
respectively, cannot receive more than $1/10$ from any action followed
by an SPNE. This makes $L$ and $x_0$ optimal in the chosen profiles (Lemma~\ref{so:upper-support}).
At the 2 earlier configurations, the positive payoff available
from $s$ and then $r$ suffices to induce entry; a best response may
choose another position (Corollary~\ref{so:further-entry}).
Thus the table specifies auxiliary histories used in the proof;
it does not assert that the entire sequence $r,s,x_0,L$ is an SPNE path.

\paragraph{Geometry used in the payoff bounds.}
The remaining position labels are
\[
\begin{aligned}
l&=2q-1/10,&U&=4q-1/10,\\
a&=1/2-2q,&b&=1/10+4q,\\
c&=1/2+2q,&d&=1/2+4q.
\end{aligned}
\]
Put
\[
\begin{aligned}
D&=(1/2-s)/2=1/5-q,&G&=(a_3-1/2)/2=1/2-4q,\\
R&=1-(a_3+r)/2=7q-1/2.
\end{aligned}
\]
The new positions in the 9-winner configuration considered below are
\[
 l,y,a,b,[x_0],c,d\qquad(L<y<U).
\]
Here $[x_0]$ is included only when the initial configuration is
$\mathcal C_5$, since there is already a candidate at $x_0$ in $\mathcal C_6$. The positions are ordered as follows:
\[l<1/10<y<s<a<b<1/2<x_0<c<a_3<d<r.\]
Figure~\ref{fig:nine-winner-family} shows the corresponding vote cells;
$v_x$ denotes the vote share of the candidate at position $x$.

\begin{figure}[!ht]
\centering
\begin{tikzpicture}[x=12.8cm,y=1cm,font=\small]
  % q=3/32 and y=1/5 are used ONLY for horizontal spacing.
  % The range arrow gives L=1/8, U=11/40 at that q.
  \draw[<->,blue!60!black] (.125,1)--(.275,1);
  \node[above,font=\scriptsize] at (.2,1.03) {$L<y<U$};
  \node[below,font=\scriptsize] at (.125,.98) {$L$};
  \node[below,font=\scriptsize] at (.275,.98) {$U$};

  \fill[blue!12] (0,-.92) rectangle (.09375,-.50);
  \fill[black!8] (.09375,-.92) rectangle (.15,-.50);
  \fill[blue!12] (.15,-.92) rectangle (.24375,-.50);
  \fill[black!8] (.24375,-.92) rectangle (.3,-.50);
  \fill[blue!12] (.3,-.92) rectangle (.4875,-.50);
  \fill[black!8] (.4875,-.92) rectangle (.53125,-.50);
  \fill[blue!12] (.53125,-.92) rectangle (1,-.50);
  \draw[black!55] (0,-.92) rectangle (1,-.50);
  \foreach \z in {
    .09375,.15,.24375,.3,.39375,.4875,.53125,.625,.71875,.8125,.90625}
    \draw[densely dashed,black!45] (\z,-.36)--(\z,-.92);
  \foreach \z/\lab in {
    .046875/q,.121875/{v_{1/10}},.196875/q,.271875/{v_s},
    .346875/q,.440625/q,.509375/{v_{1/2}},.578125/q,
    .671875/q,.765625/q,.859375/q,.953125/q}
    \node[font=\scriptsize] at (\z,-.71) {$\lab$};

  \draw (0,0)--(1,0);
  \foreach \z/\lab in {
    .1/{1/10},.2875/s,.5/{1/2},.5625/{x_0},.75/{a_3},.9375/r}
    {\fill (\z,0) circle (2pt);
     \node[above] at (\z,.08) {$\lab$};}
  \foreach \z/\lab in {
    .0875/l,.2/y,.3125/a,.475/b,.6875/c,.875/d}
    {\draw[blue!70!black,thick,fill=white] (\z,0) circle (2pt);
     \node[below] at (\z,-.08) {$\lab$};}
  \node[above] at (0,.05) {$0$};
  \node[above] at (1,.05) {$1$};

  \fill (.015,-1.32) circle (2pt);
  \node[anchor=west,font=\scriptsize] at (.03,-1.32)
    {old at $\mathcal C_6$};
  \draw[blue!70!black,thick,fill=white] (.30,-1.32) circle (2pt);
  \node[anchor=west,font=\scriptsize] at (.315,-1.32) {new candidate};
  \fill[blue!12] (.54,-1.41) rectangle (.58,-1.23);
  \node[anchor=west,font=\scriptsize] at (.595,-1.32) {winning cell};
  \fill[black!8] (.80,-1.41) rectangle (.84,-1.23);
  \node[anchor=west,font=\scriptsize] at (.855,-1.32) {losing cell};
\end{tikzpicture}
\[
v_{1/10}=\frac{y-l}{2},\qquad
v_s=\frac{a-y}{2},\qquad
v_{1/2}=\frac7{10}-7q,\qquad
0<v_{1/10},v_s,v_{1/2}<q.
\]
\caption{The 9-winner terminal configuration for
$7/80<q<1/10$ and $L<y<U$. Old and new candidates are distinguished
relative to $\mathcal C_6$; relative to $\mathcal C_5$, the candidate
at $x_0$ is also new. Dashed lines are the midpoints between consecutive distinct
candidate positions. The 9 blue cells have mass $q$; only the candidates at $1/10,s,1/2$ lose.
Horizontal spacing is shown at $q=3/32,y=1/5$, while the order and
cell formulas hold throughout the stated ranges. This configuration is used
to exclude an improvement; it is not an asserted SPNE path. The
completed configuration at $y=L$, used in the chosen subgame profile,
has 10 winners, including the candidate at $s$.}
\label{fig:nine-winner-family}
\end{figure}

The following geometric result will be used in the classification
and payoff bounds below.
\begin{lemma}\label{so:right-completion}
Use the notation above, with $7/10<a_3<4/5$.
Let $\eta$ be a history of $\Gamma$ with exactly 1 candidate at
each of $1/2,x_0,a_3,r$ and at each position in a subset of
$\{c,d\}$, and no other candidates at or to the right of $1/2$.
In every terminal extension of $\eta$ with maximal vote share $q$
in which every player entering after $\eta$ wins, there is
exactly 1 candidate at each of $x_0,c,a_3,d,r$ and no other
candidates strictly to the right of $1/2$. All 5 candidates win
with vote share $q$.
\end{lemma}
\begin{proof}
A candidate already at $c$ or $d$ has vote share $q$ at $\eta$,
and every entry affecting that candidate has capacity below $q$.
Lemma~\ref{lem:protection} therefore preserves its vote share. If $c$ is missing,
the candidate at $x_0$ has vote share $G\in(q,2q)$; if $d$ is missing,
the candidate at $r$ has vote share $R\in(q,2q)$. Every other affecting entry capacity is
below $q$. Lemma~\ref{lem:short-gaps} therefore requires exactly
1 new candidate in each missing gap, $(x_0,a_3)$ or $(a_3,r)$,
each of length $2q$.
No new candidate can enter in $(1/2,x_0)$, at $x_0$ or $r$,
or strictly to the right of $r$: these entry capacities are below $q$.
With the required gap candidates present, the vote cell at $a_3$
has length strictly below $2q$, so any copy there would have
terminal vote share below $q$.
The region from the midpoint of $1/2,x_0$ to $1$ has fixed mass
$1-(1/2+x_0)/2=5q$ and contains exactly 5 candidates.
All 5 win at $q$ by Lemma~\ref{lem:floor}.
The equalities for the vote shares at $x_0,r$ determine the new positions as
$c=1/2+2q$ and $d=2-r-2q=1/2+4q$.
\end{proof}

\begin{lemma}\label{so:upper-class}
Use the notation of this subsection, with $7/10<a_3<4/5$.
Let $\eta$ be a history of $\Gamma$ with configuration $\mathcal C_5$
and 7 players remaining, or configuration $\mathcal C_6$ and 6
players remaining, with 1 candidate at each listed position.
Let $i$ move next. Fix any action $\alpha$ of $i$ and any SPNE of
$\Gamma(\eta,\alpha)$. If $i$'s payoff in its outcome exceeds
$1/10$, the terminal configuration consists of 1 candidate at each
of the 12 positions
\[
 \{l,1/10,y,s,a,b,1/2,x_0,c,a_3,d,r\},\qquad L<y<U.
\]
Every candidate except those at $1/10,s,1/2$ wins with vote share $q$.
Player~$i$ and every player moving after $i$ enter.
\end{lemma}
\begin{proof}
Define new candidates relative to $\eta$, so the count includes
player~$i$ if that player enters. Player~$i$ wins because its payoff
is positive; every later player who enters wins by Lemma~\ref{lem:allwin}.
There are at most 7 new candidates when the configuration at $\eta$
is $\mathcal C_5$, or 6 when it is $\mathcal C_6$.
Let $k\ge1/12$ be the maximal vote share in the terminal outcome.
The gaps $(1/10,s),(a_3,r)$, and $(x_0,a_3)$ when present, have half-length $q$.
The other gaps have half-lengths $D$ at $(s,1/2)$, $G$ at $(1/2,a_3)$ in $\mathcal C_5$,
and $G-q<1/12$ at $(1/2,x_0)$ in $\mathcal C_6$.
At $\mathcal C_5$, the half-length $G<1/6\le2k$ allows at most
2 new positions in $(1/2,a_3)$ by Lemma~\ref{lem:short-gaps}.
At $\mathcal C_6$, no winning new candidate can enter in $(1/2,x_0)$,
and the same lemma allows at most 2 new positions in $(x_0,a_3)$.

\paragraph{No \(k>q\)}

If $k>q$, every entry affecting the candidate at $r$ has capacity
at most $q$ at $\eta$: the relevant bounds are $q,1-r,R/2\le q$.
Lemma~\ref{lem:protection} gives $k\ge R>1/10$.
Every entry affecting the candidate at $1/10$ has capacity at most
$1/10$, so that candidate retains vote share $1/10+q$.
Thus $k\ge1/10+q>\max\{D,G\}$, whereas every entry capacity
at $\eta$ is at most $\max\{D,G\}$.
No new candidate could attain vote share $k$, a contradiction.

\paragraph{No $k<q$.}
Apply Lemma~\ref{lem:two-sided-repair} to the old candidates at $s$ and $a_3$.
Their vote shares are $q+D=1/5$ and, respectively, $G+q$ or $2q$;
each is at least $2q$ and below $1/4\le3k$.
Their 4 adjacent gap half-lengths belong to $[q,1/6)$,
hence lie in $(k,2k)$. At least 4 players must therefore enter on each side of $1/2$.
These 8 required entries exceed the number of players remaining
at $\eta$, including player~$i$.

\paragraph{Required entry when the maximal vote share is $q$.}
Put $\varepsilon=1$ at $\mathcal C_5$ and $0$ at $\mathcal C_6$. The candidate at $1/2$ has vote share
$7/10-(6-\varepsilon)q\in((1+\varepsilon)q,(2+\varepsilon)q)$.
Its $1+\varepsilon$ adjacent gaps that can receive winning entry
have half-lengths in $(q,2q)$;
the remaining gap $(1/2,x_0)$ at $\mathcal C_6$ has entry capacity below $1/12$.
Lemma~\ref{lem:two-sided-repair} requires 2 distinct new candidates
in each of these gaps, at $a,b$ and, at $\mathcal C_5$, at $x_0,c$.
Lemma~\ref{so:right-completion} supplies any missing $c,d$ and
makes all 5 candidates at $x_0,c,a_3,d,r$ winners. For
$\mathcal C_5$, apply this geometric result after placing the
forced entries at $x_0,c$ first in the sequence of new actions.
This reordering preserves the terminal configuration and all
vote shares, so every remaining new candidate still wins.

Reducing the old candidate's vote share at $s$ to at most $q$
requires a singleton $y\ge L$ in $(1/10,s)$; its other affecting
capacities are below $q$. The candidate at $1/10$ then needs a copy or an exterior
new candidate. All remaining players must therefore enter, giving at least 9 winners:
5 on the right and 4 new candidates on the left.
A payoff above $1/10$ permits at most 9 winners. There are
therefore exactly 9, and the candidates at $1/10,s$ lose.
Hence the position $1/10$ is not copied, the new candidate in the exterior region enters at $l=2q-1/10$, and
$L<y<U$.
\end{proof}

\begingroup
\setlength{\emergencystretch}{3em}
\subsection{Completing the left positions}\label{so:alternatives-section}
We now prove the payoff bounds used at the bottom of
Table~\ref{tab:solo-backward-route}. When $L<y<U$, the next player
and the candidates specified below receive at least $1/8$ in every
SPNE. This gives a profitable deviation from an outcome with 9
winners and payoff $1/9$. At $y=L$, the corresponding bound is $1/10$;
it is used to choose SPNE profiles in Section~\ref{so:assembly-section}.
The first lemma identifies candidates who win whenever at least 1
remaining player enters and all such players win. The second proves
that the next player has an action guaranteeing a positive payoff,
so entry occurs in every SPNE.
\paragraph{Configurations and marked candidates.}\label{so:upper-notation}
Use the position labels defined in Section~\ref{so:inventory-section}.
The next 2 lemmas apply to histories of $\Gamma$ with exactly
1 candidate at each of $1/10,s,1/2,x_0,a_3,r$, at exactly one of $a,b$,
and at each position in a set $\mathcal F\subseteq\{l,y,c,d\}$, and no other candidates.
We allow $L\le y<U$. When $L<y<U$, there need not be a candidate at $y$.
Write $\ell$ for the number of positions missing among $l,y$, $j$
for the number missing among $c,d$, and $N=1+\ell+j=5-|\mathcal F|$.
There are $12-N$ old candidates. The \emph{marked candidates} are
the candidate at $x_0$ and the candidates at positions in $\mathcal F$.
When $y=L$, require a candidate at each of $y$ and $b$. The candidate at $b$ is marked as well.

\begin{lemma}\label{so:upper-completion-retention}
Let $\eta$ be a history of $\Gamma$ with either of the
\hyperref[so:upper-notation]{configurations specified above}, and use
the corresponding $N$ and marked candidates. In every terminal
extension of $\eta$ in which between 1 and $N$ players enter after
$\eta$ and every such player wins, the maximal vote share is $q$
and every marked candidate is a winner.
\end{lemma}
\begin{proof}
For a winning vote share $k<q$, Lemma~\ref{lem:floor} gives $k\ge1/12$.
The old candidate at $a$ or $b$ has vote share $D=1/5-q>1/10$.
One adjacent gap has capacity $q$; the other has capacity
$1/5-2q<1/12$, and its copy capacity is $D/2<1/12$. Since $q<2k$,
reducing this candidate's vote share to at most $k$ requires 2
distinct new candidates in the gap of capacity $q$.
The $\ell$ missing positions among $l,y$ require at least $\ell$
additional candidates to reduce the vote share of the candidate at
$1/10$ to at most $k$. With both missing, its share is $1/10+q>2q$,
which 1 new candidate cannot reduce sufficiently; with only $l$ missing
its share exceeds $q$, and with only $y$ missing it is $1/10>q$.
The old candidate at $s$ separates these entries from the gap of
capacity $q$. Missing $c,d$ require $j$ additional candidates on the
right, in regions separated by the old candidate at $a_3$. Thus $2+\ell+j=N+1$ candidates are necessary,
including copies.

Every entry capacity at $\eta$ is at most $1/10$.
The candidate already at $a$ or $b$ has vote share $D>1/10$,
and every entry affecting that candidate has capacity at most $q$.
If $k>q$, Lemma~\ref{lem:protection} preserves its share, giving
$k\ge D>1/10$. This exceeds every new candidate's attainable
vote share, a contradiction. At $q$, Lemma~\ref{so:right-completion} makes the old
candidates at $x_0,c,d$ winners whenever they are already present. Candidates already at $l,y$ have vote share $q$,
with all affecting capacities strictly below $q$: entry to the left
of $l$ has capacity below $l=2q-1/10$, its inward gap has capacity
$1/10-q$, and the 2 gaps at $y$ have capacities at most $2q-1/10$
and $3q-1/5$. Their copy capacities
are $q/2$. When $y=L$, copies of $s$ have capacity
$2q-1/10<q$, and the required $(s,b)$ singleton $x$ satisfies
$x\le L+2q=a\le x$ by the vote share bounds at $s,b$.
Thus the candidate at $b$ wins as well.
\end{proof}

\begin{lemma}\label{so:upper-shift}\label{so:upper-masked}\label{so:upper-shifted-guarantee}\label{so:h-switch}
Let $\eta$ be a history of $\Gamma$ with either of the
\hyperref[so:upper-notation]{configurations specified above}, and use
the corresponding $N$ and marked candidates. Suppose exactly $N$
players remain. In the outcome of every SPNE of $\Gamma(\eta)$, the
first remaining player and every marked candidate receive at least
$1/8$ when $L<y<U$ and at least $1/10$ when $y=L$.
\end{lemma}
\begin{proof}
Consider a sequence of entries at the missing positions: first $l$,
then $y=1/5\in(L,U)$ if $y$ is missing, then any missing $c,d$.
This leaves the missing position among $a,b$ to player~12.
Each intermediate history has the stated form, with the new candidate
among those marked in the lemma.
For $L<y<U$, filling the missing $a$ or $b$ gives 9 singleton
winners; its other neighbor, $s$ or $1/2$, strictly loses.
Player~12 can instead enter slightly closer to the candidate
already at $a$ or $b$. By Lemma~\ref{lem:deletion}, this gives 8 winners and
payoff $1/8$.
When $y=L$, filling $a$ gives 10 winners.
Player~12 therefore enters and receives at least the stated payoff
in every SPNE at the last of these histories.
Lemma~\ref{so:upper-completion-retention} makes every marked candidate
a winner in that outcome, with the same payoff as player~12.
Apply Lemma~\ref{lem:positive-chain} to the successive histories
before these prescribed actions, ending at the history where
only player~12 remains. At each history let $P_i$ be its set of
marked candidates. Each newly prescribed candidate is marked.
The retention lemma supplies condition (i), and the last-player
argument supplies condition (ii).
Thus every marked candidate receives the stated bound at each
history. The next player also receives that bound: it can take
the next prescribed action, or, if it is player~12, the final
action just described.
\end{proof}

\begin{lemma}\label{so:upper-base}\label{so:endpoint-section}
Let $7/10<a_3<4/5$, with $\mathcal C_6$ and $L$ as defined in
Section~\ref{so:inventory-section}. Consider a history $\eta$ of
$\Gamma$ with exactly 1 candidate at each position in $\mathcal C_6\cup\{L\}$,
no other candidates, and 5 players remaining.
Every SPNE of $\Gamma(\eta)$ gives each of the candidates at $L,x_0,s$
payoff $1/10$.
\end{lemma}
\begin{proof}
Consider a terminal extension of $\eta$ in which at least 1
remaining player enters and every such player wins. Let $k$ be the
maximal vote share. If $k<q$, at least 4 new candidates are needed in
$(x_0,a_3),(a_3,r)$ and separate new candidates to reduce the vote shares at
$1/10$ and $1/2$ to at most $k$, exceeding 5 remaining players.
For $k>q$, the candidate at $a_3$ keeps vote share $2q$, above every entry capacity.
At $q$, Lemma~\ref{lem:two-sided-repair} applied at $1/2$ forces $a,b$,
and Lemma~\ref{so:right-completion} forces $c,d$. Although a copy of $s$ initially has
capacity $q$, the compulsory pair $a,b$ reduces its cell to
$(a-L)/2=q$, so a copy would receive at most $q/2<q$. The candidate at $1/10$ needs a new candidate at the exterior position $l$;
its other capacities are below $q$. These 5 entries exhaust
the 5 remaining players. Every candidate except those at $1/10,1/2$
wins, including the candidates at $L,x_0,s$.
Finally, the next player can choose $b$. The resulting history
satisfies the hypotheses of Lemma~\ref{so:upper-shift}, with $y=L$,
4 players remaining and the candidate at $b$ marked.
This action therefore guarantees its mover a payoff of at least $1/10$. In every SPNE, the next player therefore enters, and the outcome has the stated 10 winners.
\end{proof}

\Needspace{9\baselineskip}
\subsection{Selecting an SPNE with further entry}\label{so:assembly-section}
We now work upwards through Table~\ref{tab:solo-backward-route}.
First use Lemma~\ref{so:upper-support} to choose a full SPNE
strategy profile in each of the 2 subgames listed there. Their outcomes give the candidate at $s$
payoff $1/10$. We then show that a candidate at $r$ wins whenever
further entry occurs, and use this to establish entry by players~5
and~4 in suitably chosen SPNE of the 2 earlier subgames.

\begin{lemma}\label{so:upper-support}
Let $7/10<a_3<4/5$. The subgames
\[
 \Gamma(1/2,1/10,a_3,r,s,x_0)\qquad\text{and}\qquad
 \Gamma(1/2,1/10,a_3,r,s)
\]
have SPNE whose first actions are $L$ and $x_0$, respectively.
Both equilibrium outcomes include entry at $L$ and
give each of the candidates at $L,x_0,s$ payoff $1/10$.
\end{lemma}
\begin{proof}
Fix either history in the statement. Suppose an action of its
next player, followed by an SPNE of the resulting subgame, gives
that player more than $1/10$. Lemma~\ref{so:upper-class} then requires
the specified 9-winner configuration, with every remaining player
entering. Let $j$ be the player who enters later at $a$ or $b$.
If a candidate is already at $x_0$ when player~$j$ moves, that player's
history satisfies the interior case of Lemma~\ref{so:h-switch}. There are $7+|\mathcal F|$ old
candidates and, because every remaining player enters in this configuration,
exactly $N=5-|\mathcal F|$ players remaining, including $j$.
The lemma gives player~$j$ at least $1/8$ in every SPNE of this subgame.
If there is no candidate at $x_0$, player~$j$ can choose $x_0$ instead.
The resulting history has $7+|\mathcal F|$ old candidates and exactly $N$
subsequent players; the mover is a marked candidate.
Lemma~\ref{so:h-switch} again guarantees that player at least $1/8$.
Both cases contradict its payoff $1/9$ in the proposed outcome.
Player~$j$ moves strictly after the initial action under consideration,
since neither $a$ nor $b$ was present at the initial history.
Its optimality is therefore required by the assumed SPNE.
This excludes a payoff above $1/10$ for the player making that
initial action.

By Lemma~\ref{so:upper-base}, every SPNE of
$\Gamma(1/2,1/10,a_3,r,s,x_0,L)$ gives the candidates at $L,x_0,s$
payoff $1/10$. Fix one such SPNE and choose arbitrary SPNE profiles
after player~7's other actions in $\Gamma(1/2,1/10,a_3,r,s,x_0)$.
Action $L$ attains the upper bound and is therefore a best response.
Use this resulting SPNE following player~6's action $x_0$ in
$\Gamma(1/2,1/10,a_3,r,s)$, and choose SPNE profiles after its other
actions. Action $x_0$ also attains $1/10$ and is a best response.
\end{proof}

\paragraph{Preserving the candidate at $r$.}
The next lemma shows that player~4, having chosen $r$, wins in
every SPNE of $\Gamma(1/2,1/10,a_3,r)$ whose outcome has further entry.
The corollary then uses player~5's action $s$ to select such an SPNE.

\begin{lemma}\label{so:lower-first}
Let $7/10<a_3<4/5$, $q=(3/2-a_3)/8$ and $r=a_3+2q$.
Fix a history $\eta$ of $\Gamma$ with exactly 4 candidates,
one at each of the positions $1/10,1/2,a_3,r$. In every terminal
extension of $\eta$ in which between 1 and 8 players enter after
$\eta$ and every such player wins, the old candidate at $r$ is a winner.
\end{lemma}
\begin{proof}
Each new candidate has the winning vote share $k$, which is at least $1/12$
since at most 12 players have entered.

We first show that $k\le q$ in every such terminal extension. If $k>q$, the
affecting capacities $q,1-r,R/2$ leave the candidate at $r$ with
unchanged vote share $R$, so $k\ge R>1/10$.
For $k\ge3/20$, the candidate at $a_3$ also keeps its vote share: its capacities
$G,q,(G+q)/2$ are below $3/20$, while its vote share
$G+q$ exceeds $1/5$, the upper bound on all entry capacities.
For $1/10<k<3/20$, Lemma~\ref{lem:two-sided-repair} at the extreme position
$1/10$ excludes entry: its vote share $3/10$ lies in $(2k,3k)$, its interior
capacity $1/5$ lies in $(k,2k)$, and its exterior capacity is below $k$.
Thus $k>q$ is impossible.

For $k<q$, Lemma~\ref{lem:two-sided-repair} at $a_3$ forces
strict pairs in $(1/2,a_3),(a_3,r)$: $2k<G+q<3k$ and $G,q\in(k,2k)$.
Together with the 5 new candidates at or left of $1/2$ from
Lemma~\ref{lem:two-anchor-region}, these exceed 8 decisions.
At $k=q$, the copy and exterior capacities of $r$ are below $q$,
so $(a_3,r)$ contains a singleton $z\ge d=1/2+4q$.
Let $\varepsilon=0,1$ indicate whether $(1/2,a_3)$ contains its strict pair.
The vote cell at position $a_3$, shared by all candidates there,
has length
\[
 M=(z-1/2)/2-\varepsilon q\in[(2-\varepsilon)q,(3-\varepsilon)q).
\]
If $a_3$ is not copied, its candidate's vote share cannot exceed $q$;
if $a_3$ is copied, its cell mass is an integer multiple of $q$. Thus $M=(2-\varepsilon)q$, giving $z=d$
and making the old candidate at $r$ a winner.
\end{proof}

\begin{corollary}[Further entry]\label{so:further-entry}
For every $a_3\in[1/2,4/5]$, the subgame $\Gamma(1/2,1/10,a_3)$ has an
SPNE in which at least 1 of players~4--12 enters.
\end{corollary}
\begin{proof}
Lemma~\ref{so:direct-entry} proves the claim for
$a_3\in[1/2,7/10]\cup\{4/5\}$. Suppose now $7/10<a_3<4/5$.
In $\Gamma(1/2,1/10,a_3,r)$, choose the SPNE from
Lemma~\ref{so:upper-support} following the fifth-player action $s$,
which gives that player payoff $1/10$. Choose full SPNE profiles
after all other fifth-player actions. Every best response by player~5 to these profiles gives it at least
$1/10$ and is therefore entry. Lemma~\ref{so:lower-first} then makes
player~4, at $r$, a winner with a positive payoff.
Use this SPNE after player~4's action $r$ and choose SPNE profiles
after its other actions. A best response by player~4 to these
profiles also gives it a positive payoff and is therefore entry,
as required.
\end{proof}

\endgroup

Corollary~\ref{so:further-entry} completes Case 1.
Section~\ref{sec:positive-continuation} checks that the profiles
can be chosen compatibly when more than one configuration is
possible after the same action.

\section{Proof of the main result}\label{sec:positive-continuation}
We now complete the proof using the 2 requirements in
Section~\ref{sec:mechanism}. We first verify that an SPNE is
available in every subgame, then choose compatible profiles after
all actions of player~3. We finally choose best responses for
players~3, 2, and 1, in that order.

\paragraph{Existence of subgame equilibria.}
Every subgame has a pure SPNE. To see this, induct on the number
of remaining players. With no players remaining, the empty
strategy profile is an SPNE. At any earlier decision, the
induction hypothesis supplies an SPNE strategy profile after
each available action. Choose one such profile for each action.
All terminal payoffs belong to the finite set
$\{-1,0,1/12,1/11,\ldots,1/2,1\}$, so the resulting set of
payoffs available to the current player is nonempty and finite.
Some action therefore attains its maximum. Combining that action
with the chosen subgame profiles gives an SPNE at the current
decision. Histories following distinct actions form disjoint
subtrees, so the strategy specifications are consistent.
Set-theoretic choice permits these selections over the continuum
of actions. This proves existence in every subgame, including
the full game $\Gamma$.

\begin{proposition}\label{prop:positive-child}
The subgame $\Gamma(1/2,1/10)$ has a pure SPNE in which player~2
receives a payoff of at least $1/10$.
\end{proposition}

\begin{proof}[Proof of Proposition~\ref{prop:positive-child}]
We choose one full SPNE strategy profile of
$\Gamma(1/2,1/10,a_3)$ for each available action $a_3$.
Table~\ref{tab:case-coverage} records the results used to make
these choices. It uses the positions and parameter ranges from
Table~\ref{tab:improvement-families} and concerns $a_3\in[0,1]$;
action $\OUT$ gives player~3 payoff zero.
``Every'' means that the indicated outcomes are excluded in every
SPNE of the subgame. ``Choice'' means that an SPNE avoiding those
outcomes can be selected.
\Needspace{20\baselineskip}
\begin{longtable}{@{}>{\raggedright\arraybackslash}p{0.25\textwidth}>{\raggedright\arraybackslash}p{0.25\textwidth}>{\raggedright\arraybackslash}p{0.12\textwidth}>{\raggedright\arraybackslash}p{\dimexpr0.38\textwidth-6\tabcolsep\relax}@{}}
\caption{Exclusion of the harmful outcomes in Table~\ref{tab:improvement-families}.}\label{tab:case-coverage}\\
\toprule \textbf{Configuration} & \textbf{Third action} & \textbf{Scope} & \textbf{Argument}\\\midrule\endfirsthead
\toprule \textbf{Configuration} & \textbf{Third action} & \textbf{Scope} & \textbf{Argument}\\\midrule\endhead
Case 1 & $1/2\le a_3\le7/10$ & Every & Lemma~\ref{so:direct-entry}.\\
Case 1 & $7/10<a_3<4/5$ & Choice & Corollary~\ref{so:further-entry}.\\
Cases 1 and 3 & $a_3=4/5$ & Choice & Proposition~\ref{prop:case-three}, including its further-entry conclusion.\\
Case 2 & $a_3\in\{3/20,9/20,1/2\}$ & Every & Section~\ref{sec:copied-families}.\\
Case 3 at $V=7/40$ & $a_3\in\{3/20,9/20,$\newline $17/20\}$ & Every & Proposition~\ref{bd:excluded}.\\
Case 3: distinct positions, $V>7/40$ & $a_3\in\{a,b,r,s\}$ & Every & Proposition~\ref{sf:main}.\\
Case 3: copied rightmost position, $V>7/40$ & $a_3\in\{a,b,r\}$ & Every & Lemma~\ref{lem:case-three-copies}.\\
Case 4: distinct positions & $a_3\in\{a,b,s\}$ & Every & Corollary~\ref{cor:five-entry-order}.\\
Case 4: distinct positions & $a_3\in\{r,t\}$ & Choice & Proposition~\ref{prop:case-four}.\\
Case 4: coincident positions & Any position listed for this case & Every & Section~\ref{sec:case-four-copies}.\\
\bottomrule
\end{longtable}

\medskip
\noindent\textit{Compatible choices of subgame profiles.}
If $a_3\in[0,1]$, player~3 is the first new candidate.
Corollary~\ref{cor:five-entry-order} therefore implies that a
strict Case 4 outcome can occur only if $a_3=r$ or $a_3=t$.
Since $3/20<V<1/6$, these actions lie in
$(1/2,3/5)\cup(5/6,9/10)$. We make the following choices:
\begin{enumerate}[label=(\roman*),nosep]
\item At $a_3=9/10$, choose any SPNE of $\mathcal S$.
By Proposition~\ref{prop:attainment}, its outcome gives players~2
and~3 the same payoff $\rho\ge1/10$.
\item For $7/10<a_3<4/5$, choose an SPNE with further entry
as provided by Corollary~\ref{so:further-entry}.
\item At $a_3=4/5$, use Proposition~\ref{prop:case-three} to
choose an SPNE avoiding Case 3 and including further entry.
\item For $a_3\in(1/2,3/5)\cup(5/6,9/10)$, choose an SPNE
avoiding Case 4 as provided by Proposition~\ref{prop:case-four}.
\item After every other action, including $\OUT$, choose any SPNE.
\end{enumerate}

The action sets in (i)--(iv) are pairwise disjoint. For
$a_3\in[0,1]$, all exclusions hold simultaneously as follows.
Case 2 and coincident-position outcomes in Case 4 are excluded
in every SPNE. Case 3 is excluded in every SPNE when $a_3\ne4/5$;
choice (iii) handles $a_3=4/5$. Case 1 is excluded in every SPNE
on $[1/2,7/10]$, and choices (ii)--(iii) exclude it on
$(7/10,4/5]$ by further entry. Finally, the entry-order
restriction above excludes strict Case 4 outcomes outside the
intervals in (iv), while (iv) excludes them within those intervals.
For example, at $a_3=17/20$ the profile chosen under (iv) also
avoids Case 3, since that case is excluded there in every SPNE.

Proposition~\ref{prop:improvement-classification} now implies
that, in every chosen profile, a payoff above $1/10$ for player~3
makes player~2 a winner as well; for $a_3=\OUT$ the premise is
false. Both winning players receive the same payoff. Thus these profiles satisfy both requirements in
Section~\ref{sec:mechanism}.

\medskip
\noindent\textit{Player 3's best response.}
With these subgame profiles fixed, the payoffs available to
player~3 form a nonempty finite set, so a best response exists.
Choose $9/10$ whenever it is a best response. If this action is
chosen, player~2 receives $\rho$. Otherwise, the chosen action
gives player~3 strictly more than $\rho\ge1/10$, and the preceding
implication gives player~2 the same payoff. Combining this best
response with the full subgame profiles gives an SPNE of
$\Gamma(1/2,1/10)$ in which player~2 receives at least $1/10$.
\end{proof}

\Needspace{6\baselineskip}
\begin{proof}[Proof of Theorem~\ref{thm:main}]
Fix the SPNE of $\Gamma(1/2,1/10)$ from
Proposition~\ref{prop:positive-child}. For every other action
$a_2\in[0,1]\cup\{\OUT\}$ of player~2, choose an SPNE of
$\Gamma(1/2,a_2)$. Against these fixed subgame profiles, action
$1/10$ gives player~2 at least $1/10$. A best response therefore
gives player~2 at least $1/10$ and differs from $\OUT$.
Combining that action with the chosen profiles yields an SPNE
of $\Gamma(1/2)$ in which player~2 enters.

Retain this SPNE after player~1's action $1/2$ and choose an
SPNE of $\Gamma(a_1)$ after every other first action $a_1$.
A best response by player~1 exists by the same finite-payoff
argument. Together with these full subgame profiles, it gives
an SPNE of $\Gamma$. If player~1 chooses $1/2$, player~2 enters;
if player~1 chooses another action, the first action differs
from that in $a^{\mathrm{can}}$. In either case, the resulting
outcome differs from $a^{\mathrm{can}}$, proving the theorem.
\end{proof}

Some subgame profiles and the maximizing actions of players~1--3
remain implicit in this existence proof. Each selected profile specifies
strategies at every history in its subgame, including histories off the
equilibrium path. The SPNE of $\Gamma(1/2,1/10)$ gives player~2 a positive
payoff; the proof does not require $(1/2,1/10)$ to be the first 2 actions
on the resulting full-game equilibrium path.

\section{Conclusion}\label{sec:conclusion}
The Osborne--Kats conjecture predicts that only the first and last
players enter, both at the median. With 12 players and a uniform
electorate, we establish a pure SPNE with a different outcome.
Starting from the first 2 entries $(1/2,1/10)$ proposed by de Vries,
we establish the existence of a subgame SPNE giving player~2 a positive
payoff and extend it to the full game. This refutes the conjectured uniqueness. Whether
the canonical outcome can coexist with this equilibrium, and whether
fewer players suffice, remain open.

The proof uses the incentives of later players to enter as winners.
Geometric restrictions on those entries can force candidates who
entered earlier to win. These restrictions allow the chosen
subgame strategy profiles to support entry away from the median. A profitable move that reduces the number of tied winners must
also withstand the remaining players' responses. These 2 observations
underlie the retention and deviation arguments.

The construction uses exact equalities between vote shares at distinct positions. Perturbing the voter
distribution can break these ties and change payoffs discontinuously,
even when vote shares change continuously. An extension would need to
adjust candidate positions and recheck both the tie equations and the
entry bounds. For more than 12 players, additional decisions alter
the possible responses to earlier deviations. The arguments based on
the number of remaining players must therefore be revisited.

\section*{Lean formalization}
Theorem~\ref{thm:main} has also been formalized in Lean~4.
The Lean source code and instructions for checking the formal proof
are available at \url{https://www.dagaev.com/publications}.

\section*{Acknowledgments}
The author is grateful to Martin Osborne for maintaining
a comprehensive public record of the conjecture and for
his valuable comments that helped improve the manuscript.

\section*{Funding}
Support from the Basic Research Program of HSE University is gratefully acknowledged (HSE-BR-2025-082). 

\par\bigskip\noindent\begin{minipage}{\textwidth}
\section*{Declaration of generative AI and AI-assisted technologies}
OpenAI's GPT-6 Astra was used throughout manuscript preparation to assist with the development of mathematical arguments and the organization and editing of the text and LaTeX source. The author takes full responsibility for the final content of this work.
\end{minipage}\par

\end{document}